\documentclass[a4paper,twocolumn]{revtex4}
\usepackage{amsmath}
\usepackage{amsfonts}
\usepackage{amssymb}
\usepackage{graphicx}
\usepackage{subscript}
\usepackage{siunitx}
\usepackage{xcolor}
\usepackage{gensymb}

\begin{document}

\title{How patchiness controls the properties of chain-like assemblies of colloidal platelets}

\author{Carina Karner}
\email{carina.karner@univie.ac.at}
\affiliation{Faculty of Physics, University of Vienna, Boltzmanngasse 5, A-1090, Vienna, Austria}

\author{Christoph Dellago}
\affiliation{Faculty of Physics, University of Vienna, Boltzmanngasse 5, A-1090, Vienna, Austria}

\author{Emanuela Bianchi}
\email{emanuela.bianchi@tuwien.ac.at}
\affiliation{Institut f{\"u}r Theoretische Physik, TU Wien, Wiedner Hauptstra{\ss}e 8-10, A-1040 Wien, Austria }
\affiliation{CNR-ISC, Uos Sapienza, Piazzale A. Moro 2, 00185 Roma, Italy}

\date{\today}

\begin{abstract} 
Patchy colloidal platelets with convex, non-spherical shapes have been realized with different materials at length scales ranging from nanometers to microns. While the assembly of these hard shapes tends to maximize edge-to-edge contacts, as soon as a directional attraction is added -- by means of, $e.g.$, specific ligands along the particle edges -- a competition between shape and bonding anisotropy sets in, giving rise to a complex assembly scenario. We focus here on a two-dimensional system of patchy rhombi, $i.e.$, colloidal platelets with a regular rhombic shape decorated with bonding sites along their perimeter. Specifically, we consider rhombi with two patches, placed on either opposite or adjacent edges. While for the first particle class only chains can form, for the latter we observe the emergence of either chains or loops, depending on the system parameters. According to the patch positioning -- classified in terms of different classes, topologies and distances from the edge center -- we are able to characterize the emerging chain-like assemblies in terms of length, packing abilities, flexibility properties and nematic ordering.

\end{abstract}

\maketitle

\section{Introduction}
The interest in the self-assembly of colloidal chains stems from a variety of different applications. On one hand, colloidal chains can be seen as larger-scale version of linear polymers and, thus, they can serve as model systems to investigate the chain dynamics: the advantage of colloidal chains with respect to polymers and biopolymers is that they can be observed on a single-particle level in real space with optical techniques~\cite{Liu2010, Rao2012}. As, for instance, the flexibility of colloidal chains can be tuned from the rigid to the semi-flexible down to the flexible regime~\cite{Rao2012}, its effect on the dynamics in the response to an external field can be investigated~\cite{Kuei2017}, thus providing insights on natural phenomena -- such as flagellar motion~\cite{Zoettl2019} or conformational transitions in polymer systems~\cite{Kirchenbuechler2014} -- as well as on how to fabricate and optimize microfluidic devices~\cite{Vilfan2010}. On the other hand, colloidal chains are {\it per se} applicable, e.g., as microscale detectors~\cite{Su2016}, responsive materials with optical properties~\cite{Hu2011} and wavy arrays of self-assembled colloidal fibres for specific functionalization or coating~\cite{Demortiere2014}.

The emergence of linear assemblies usually relies on directional interactions. Anisotropic interactions can be introduced, for instance, through the application of external electric or magnetic fields~\cite{Bharti2015} or via the design of interaction patterns generated by well-defined bonding sites, also referred to as ``patches''~\cite{patchyrevexp,patchyrevtheo,newreview}. In the first case, external fields induce a dipole moment that leads to a preferred assembly along the direction of the applied field. As the resulting chains disassociate to individual particles when the fields are switched off, an extra experimental step is needed to permanently link the connected particles~\cite{Rao2012}. The field-directed assembly can also be used to make structures from non-spherical particles; in this case, the assembly is governed by two factors: the polarization of particles and the entropic interactions related to the particle shape~\cite{Herlihy2008,Yanai2013}. In contrast, patchy colloids can provide directional interactions even in the absence of an external field~\cite{patchyrevexp,patchyrevtheo,newreview}. Colloidal patchy polymers have been for instance studied to investigate and reproduce the folding of proteins at micrometer scale~\cite{Coluzza2013}. We note that, solid patches are usually used but liquid patches are also a viable way to induce one-dimensional assembly~\cite{Zhao2019}. The two approaches -- patches and external fields -- can be combined, thus producing interesting responsive materials for application in, e.g., micro-robotics~\cite{Yan2013,Shah2015}.

Here we study the formation of chain-like assemblies emerging in two-dimensional systems of non-spherical patchy particles. Colloidal platelets of different shapes can be realized experimentally at the nano- up to the micro-scale: polygonal truncated silica pyramids, lanthanide fluoride nanocrystals and DNA-origami of several shapes are just a few examples~\cite{dna_origami,platelets_review}. Additional directionality in bonding can be imparted to the systems by, $e.g.$, covering the colloidal edges with ligands~\cite{patchy_platelets} or immersing them in a liquid crystal medium~\cite{rhombi_liqcry}. While hard shapes assemble by maximizing their edge-to-edge contacts, the additional bonding pattern induced by the patches favors configurations where the number of bonds can be maximized. 

In our investigation, we consider rhombi platelets decorated with two attractive patches in various geometries: patches can be placed either on opposite or adjacent edges. Within these two big classes of systems, we explore the assembly scenario resulting from different patch positioning, focusing on those where chain-like assembly prevails. 

The paper is organized as follows: in Sec.~\ref{sec:model&methods} we describe the patchy rhombi model (Sec.~\ref{sec:model}) and provide the details of our two-dimensional Monte Carlo simulations (Sec.~\ref{sec:methods}), in Sec.~\ref{sec:results} we discuss our results, first for systems where only chains can form (Sec.~\ref{sec:pl}) and then for systems where chains compete with loops and micelles (Sec.~\ref{sec:mamo}). Finally in Sec.~\ref{sec:conclusions} we draw our conclusions. 

\section{Model and Methods}\label{sec:model&methods}

\subsection{Particle model}\label{sec:model}
\begin{figure}[h]
\centering
\includegraphics[width=0.3\textwidth]{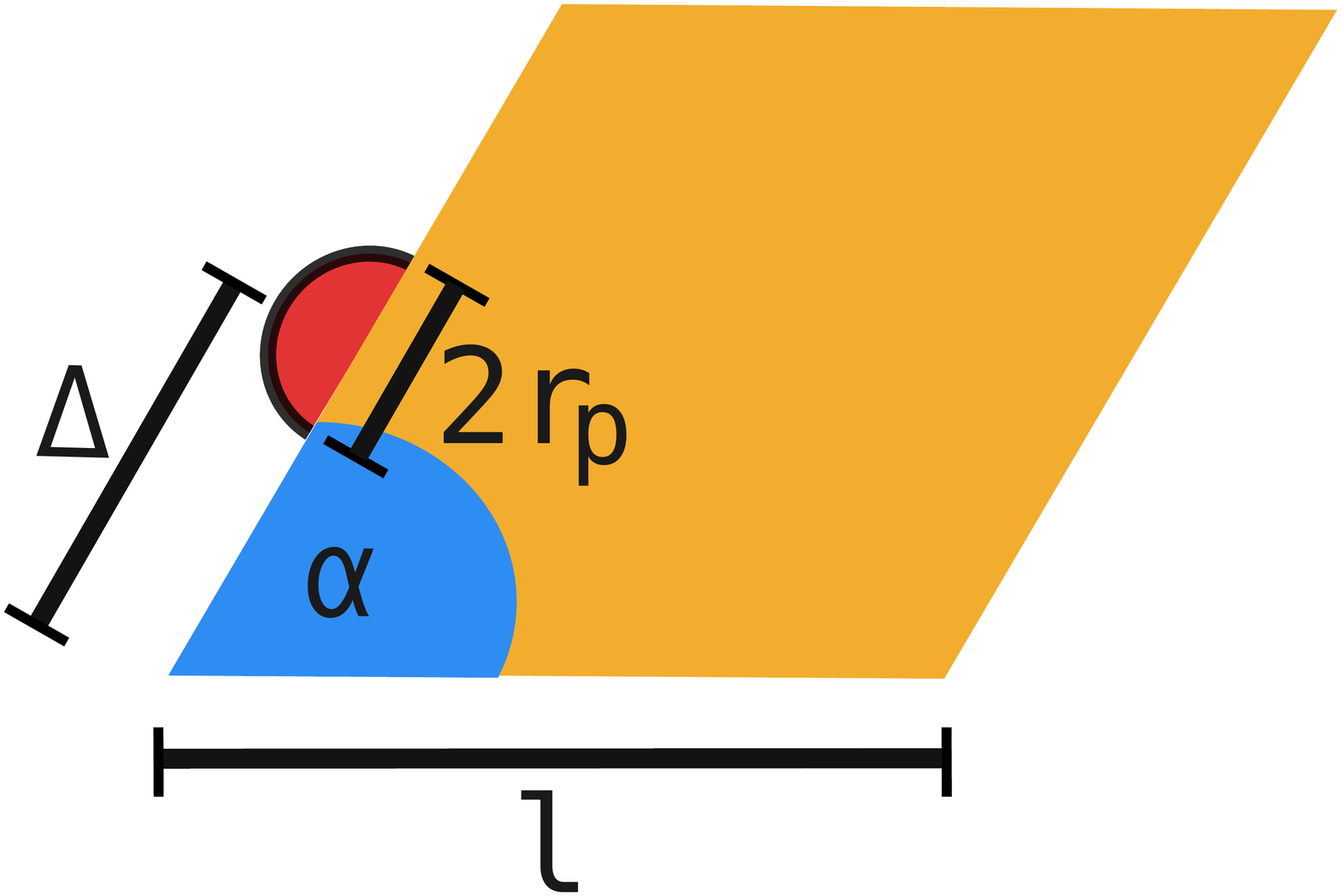}
\caption{Rhombic platelet with one patch: the edge size $l$ sets the unit length, the opening angle $\alpha=60^{\degree}$ for a regular rhombus, the patch size $2r_p=0.1$, as in Table~\ref{table:geom}. The patch position is determined by the parameter $\Delta$, as labelled. 
}
\label{fig:sketch}
\end{figure}

Our particles are regular hard rhombi (see the sketch in Fig.~\ref{fig:sketch}) with two attractive square-well interaction sites, denoted as patches in the following and placed on different edges. In general the interaction potential between two hard particles $i$ and $j$ is given by
\[ U(\vec{r}_{ij}, \Omega_{i}, \Omega_{j})  =
  \begin{cases}
    0     & \quad \text{if  $i$ and $j$ do not overlap}\\
    \infty  & \quad \text{if $i$ and $j$ do overlap}.\\
  \end{cases}
\]
with $\vec{r}_{ij}$ as the center-to-center vector, and $\Omega_{i}$ and $\Omega_{j}$ as particle orientations. 
To determine the overlaps between the rhombi we employ the separating axis theorem for convex polygons, detailed in Ref.~\cite{Golshtein_1996}. The patch-patch potential is a square-well defined by
\[ W(p_{ij})  =
  \begin{cases}
    - \epsilon     & \quad \text{if}\quad p_{ij}< 2r_{p}\\
    0 & \quad  \text{if}\quad p_{ij} \geq 2r_{p}, \\
  \end{cases}
\]
where $p_{ij}$ is the patch-patch vector, $2r_{p}$ is the patch diameter and $\epsilon$ denotes the patch interaction strength. A patchy rhombi model of this kind was first introduced in Ref.~\cite{Whitelam2012}, with four attractive patches placed in the center of the edges. 

Regular rhombi with two patches can be classified in three configuration types, as there are three ways  to distribute two patches on four edges: first, patches can be placed on opposite edges -- a configuration referred to as parallel (pl) --  secondly, patches can be placed to enclose the big angle -- a configuration denoted as manta (ma) -- and thirdly, patches can be placed to enclose the small angle -- a configuration denoted as mouse (mo). For a sketch of pl-class see the top of Fig.~\ref{fig:pl_clusters}, for ma- and mo- systems see the top of Fig.~\ref{fig:mamo_clusters}. In each particle class, patches can be placed anywhere on their respective edges, resulting in an -- in principle -- infinite number of possible patchy rhombi. To study these systems methodically and give instructive design directions we introduce symmetric and asymmetric patch topologies. Patch topologies -- or movement patterns -- prescribe how to move the patches with respect to each other when scanning the parameter space.  In the symmetric (s) topology, patches are always placed symmetrical with respect to their enclosing vertex (ma and mo systems) or such that patches on opposing edges sit exactly opposite to each other (pl systems).  In contrast, in the asymmetric (as) topology, patches are placed asymmetrically with respect to the enclosing angle (ma- and mo systems) or such that the patch positions are mirrored with respect to the edge center (pl systems).  Note that it suffices to state the relative position $\Delta$ of only one of the two patches with respect to the reference vertex, as the second one is automatically defined through the choice of topology (see  Fig.~\ref{fig:pl_clusters} for pl-systems, and Fig.~\ref{fig:mamo_clusters}). With these definitions a two-patch system is fully defined through its patch configuration (pl, ma or mo), its topology (s or as) and its relative position on the edge ($\Delta$).  It is important to note that in the edge-center, $i.e.$, at $\Delta=0.5$, the s- and as-topology collapse into the center topology and the respective systems are denoted as pl-, ma- and mo-center systems. A summary of the used particle parameters can be found in Table~\ref{table:geom} and Fig.~\ref{fig:sketch}.
We note that, for pl-systems we show results up to $\epsilon=-8.2 k_BT$, where particles mostly belong to chains, while for ma- and mo-systems we show only $\epsilon=-10.2 k_BT$, as at lower attraction strengths the competition with other assembly products plays a non-negligible role. For an extensive discussion about the $\epsilon$-dependent behavior of ma- and mo-systems we refer to Ref.~\cite{micelles}.

\begin{table}[h]
\begin{center}
\begin{tabular}{ |l|l|l| } 
\hline
 \bf{parameter} & \bf{symbol} & \bf{value} \\
 \hline
 angle & $\alpha$ & $60\degree$ \\  
 \hline
 side length & $l$ & 1.0 \\ 
 \hline
 patch radius & $r_{p}$  & 0.05 \\
 \hline
 interaction strength & $\epsilon$ & -4.2 -- -10.2 \\
 \hline
 patch position  & $\Delta$ & 0.2 -- 0.8 \\
 \hline
\end{tabular}
\caption{Single particle parameters. See Fig.~\ref{fig:sketch} for a graphical representation of the edge length $l$, the opening angle $\alpha$, the patch radius $r_p$ and the patch position parameter $\Delta$.}
\label{table:geom}
\end{center}
\end{table}

\subsection{Simulation details}\label{sec:methods}
We model the adsorption of the platelets on a surface with grand canonical ($\mu VT$ ensemble) simulations with single particle rotation and translation moves and particle insertion and deletion.  Cluster moves are added to avoid kinetic traps~\cite{Whitelam2007, Whitelam2010}. All systems are initially equilibrated for $3\times10^5$ MC-sweeps at very low packing fraction of $\phi\approx 0.05$ with a fixed chemical potential $\mu_{eq}$. After equilibration, we increase the chemical potential to $\mu^{*}$ to observe the assembly.  We performed 8 simulations runs per system and interaction strength $\epsilon$. We note that for pl-center systems we collected twice the statistics as we run both pl-s and pl-as with $\Delta=0.5$; most of the analysis is performed by considering these two data sets as independent, but of course they yield to the same results. For ma- and mo-systems, we run the simulations for about  $\approx 3\times 10^6 - 5.0\times 10^6$ MC-sweeps before collecting statistics.  Pl-systems assembled faster and the total duration of the runs is $\approx 1\times 10^6$ MC-sweeps. The system parameters for all simulations are are given in Table~\ref{table:system_param}.

\begin{table}[h]
    \begin{center}
    \begin{tabular}{|l|l|l|}
        \hline
        \bf{system parameter} &  \bf{symbol} & \bf{value} \\
        \hline
         area of simulation box & A & $1000 \cdot \sin{(60\degree)}$ \\
         \hline
         box width & $L_{x} $ & $\sqrt{1000}$ \\
         \hline
         box height & $L_{y}$ &  $\sqrt{1000}
         \cdot\sin{(60\degree)}$ \\
         \hline
         chemical potential eq. & $\mu_{eq}$ & 0.1 \\
         \hline
         chemical potential & $\mu^{*}$ & 0.3 \\
         \hline
         Boltzmann constant  & $k_{\rm B}$ & 1 \\
         \hline
         Temperature & $T$ & 0.1\\ 
         \hline
    \end{tabular}
    \caption{The system parameters used in all simulations.}
    \label{table:system_param}
    \end{center}
\end{table}

\section{Results}\label{sec:results}

In general, on varying the patch positioning $\Delta$ and the patch-patch energy strength $\epsilon$, two-patch rhombi yield three different classes of self-assembly products: chains, loops and micelles, where micelles can be defined as minimal loops. While pl-systems form chains for any $\Delta$ and $\epsilon$ value, ma- and mo-systems form both chains and loops. In our discussion we mostly focus on characterizing and comparing the emerging chains in different systems and we postpone the analysis of the emerging loops and micelles in ma- and mo-systems to Ref.~\cite{micelles}.  The different assembly scenarios result from the interplay between steric constraints and patchiness. By construction, our patchy rhombi can form only one bond per edge, for a total of two bonds per particle. A pair of such rhombi can bind in two possible orientations: parallel (p) -- $i.e.$, with their long axes oriented parallel to each other -- and non-parallel (np) -- $i.e.$, with they long axes in an arrowhead orientation. It is worth noting already at this stage that chains contain only p-bonds, micelles contain only np-bonds, while loops consist of both. 

\subsection{Chains in pl-systems}\label{sec:pl}

\begin{figure*}
\centering
\includegraphics[width=\textwidth]{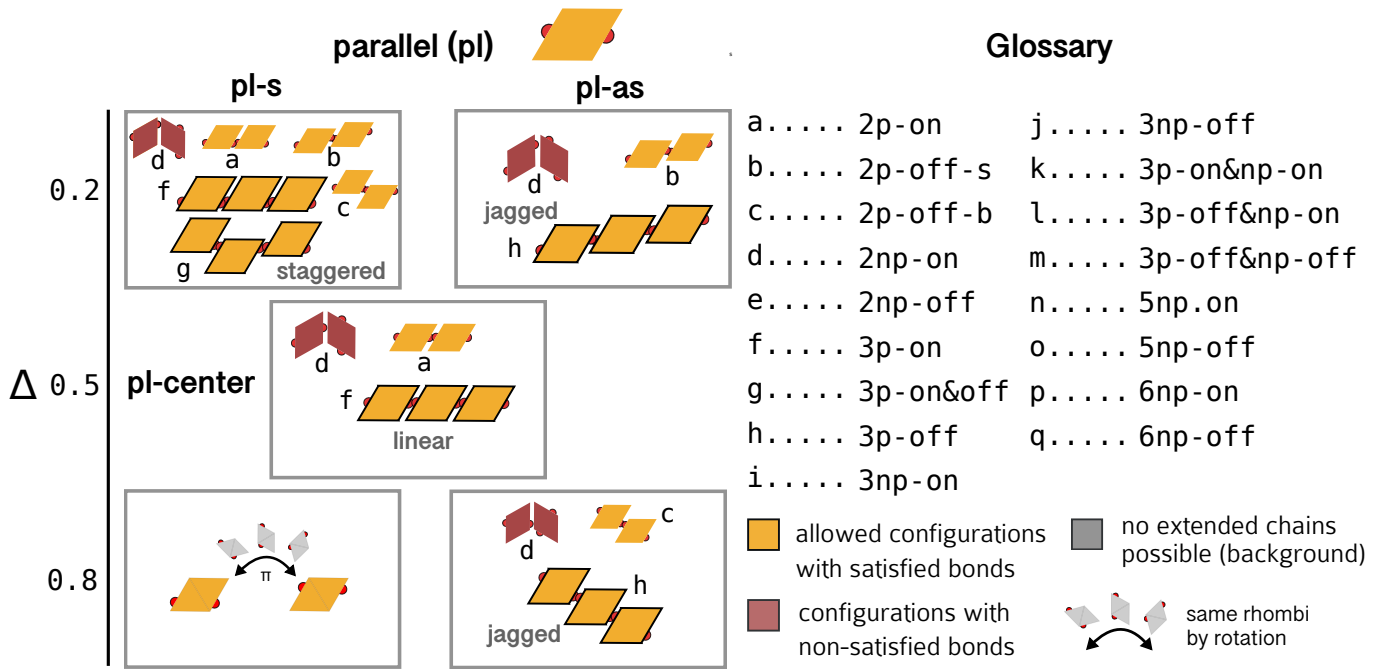}
\caption{Right: small cluster analysis for parallel (pl) systems; left column: symmetric topology (pl-s), right column: asymmetric topology (pl-as), as labeled. The row corresponds to $\Delta = 0.2$ (top), 0.5 (center) and 0.8 (bottom), as labeled. Note that at $\Delta = 0.5$, both pl-s and pl-as collapse to pl-center. Clusters with fully bonded rhombi are colored in orange, while clusters with non-satisfied bonds are colored in burgundy. Note that pl-s with $\Delta = 0.2$ and pl-s with $\Delta = 0.8$ yield the same system by rotation of $\pi$, while pl-as are not symmetric with respect to $\Delta=0.5$. Left: glossary of the labeling of the different clusters. Dimers with on-edge parallel bonds (2p-on) are labeled with a, dimers with off-edge parallel bonds (2p-off) are labeled with either b (off-edge bonds closer to the small angle, 2p-off-s) or c (off-edge bonds closer to the big angle, 2p-off-b), dimers with non-parallel bonds (2-np) are labeled with either d (on-edge bonds, 2-np-on) or e (off-edge bonds, 2-np-off). The same logic applies to trimers and bigger clusters.}
\label{fig:pl_clusters}
\end{figure*}

\begin{figure*}
\begin{center}
\includegraphics[width=\textwidth]{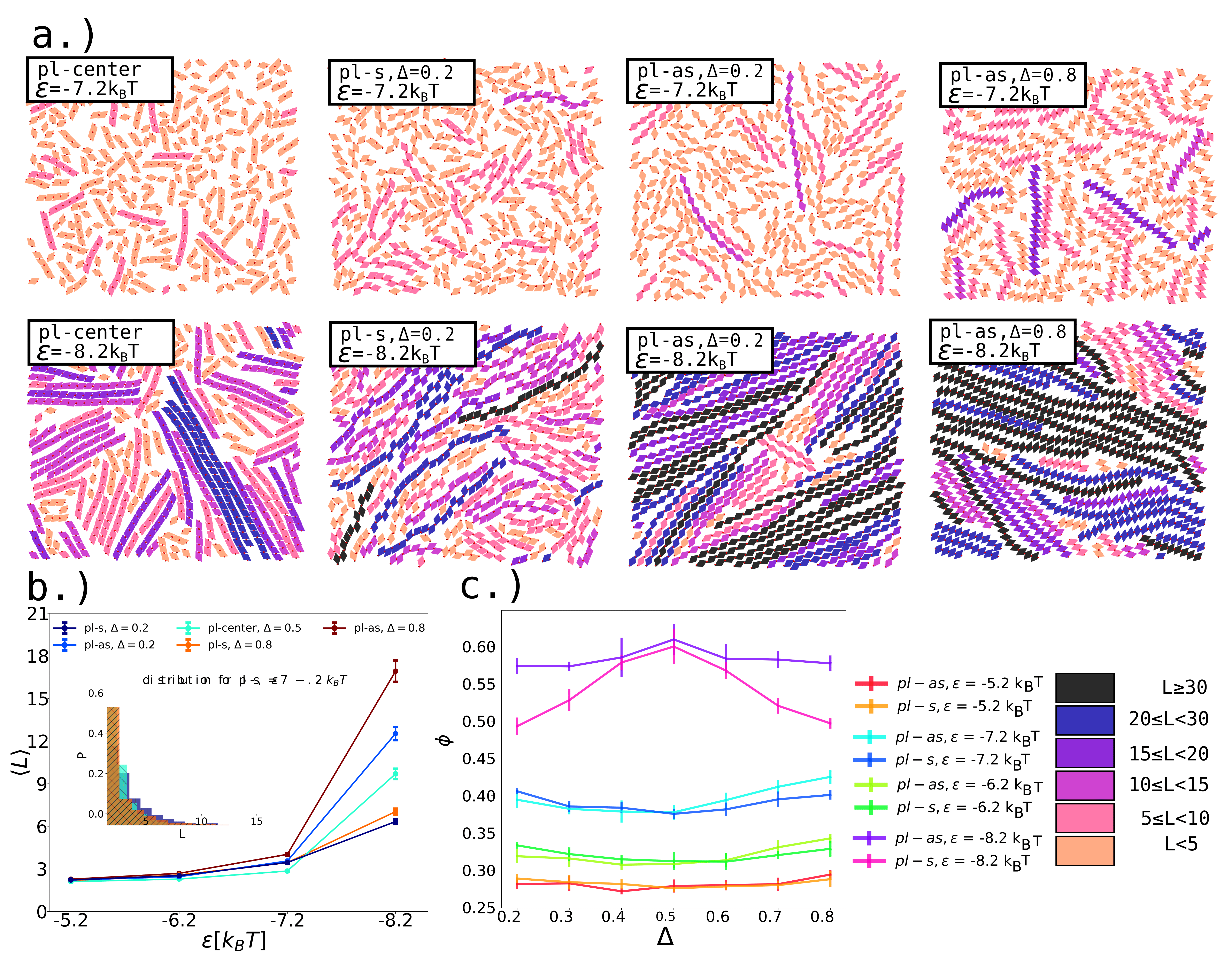}
\caption{\textbf{a.)} Snapshots of pl-center (first column, linear chains), pl-s  (second column, staggered chains) and pl-as (third and fourth columns, jagged chains) colored according to the chain length (see color scale at the bottom) for interactions strengths $\epsilon = -7.2 k_{B}T$ (top row) and $\epsilon = -8.2 k_{B}T$ (bottom row). While pl-s-rhombi with $\Delta=0.2$ and $\Delta=0.8$ are the same by rotation of $\pi$, pl-as-rhombi form two types of jagged chains: for $\Delta=0.2$ particles bond off-edge with the bonds closer to the small rhombi angle, while for $\Delta=0.8$ particles bond off-edge with the bonds closer to the big rhombi angle.  \textbf{b.)} Average chain lengths $\langle L \rangle$ of pl-systems (as labeled) as function of interaction strength $\epsilon$. Inset: distribution of chain length for $\epsilon = -7.2k_{B}T$  for all pl-systems, as labeled \textbf{c.)} Packing fraction $\phi$ of pl-systems as function of the patch position $\Delta$ at different interaction strengths $\epsilon$.}
\label{fig:pl-clength}
\end{center} 
\end{figure*}

We start our discussion with the analysis of pl-systems.  As anticipated, all pl-systems, namely pl-center, pl-s and pl-as, form p-bonded chains for all choices of $\Delta$ and $\epsilon$. This observation can already be deduced from the small cluster analysis reported in Fig.~\ref{fig:pl_clusters}: this analysis determines which clusters can form depending on system type, topology and patch position. In pl-systems the patch positioning is not compatible with np-bonds (see dimers referred to as d in all panels of Fig.~\ref{fig:pl_clusters}) and thus pl-rhombi can assemble only in chains (see configurations referred to as f and g), with either on-edge (configurations a) or off-edge (configurations b) p-bonds.
In particular, chains in the pl-center topology consist exclusively of on-edge bonds, pl-as have only off-edge bonds and pl-s chains can have both.
These bonding constraints lead to chain types that differ from each other in appearance as well as in physical properties. The on-edge bonds of pl-center result in \textit{linear} chains, the off-edge bonds of pl-as leads to \textit{jagged} chains and the mix of on-edge and off-edge bonds in pl-s yields \textit{staggered} chains (see Fig.~\ref{fig:pl_clusters} and Fig.~\ref{fig:pl-clength}a for simulation snapshots). It is important to note that there are two different types of jagged chains: for $\Delta<0.5$ pl-as-rhombi bond off-edge with the bonds closer to the small rhombi angle, while for $\Delta>0.5$ pl-as-rhombi bond off-edge with the bonds closer to the big rhombi angle. In contrast, pl-s-rhombi are the same by rotation of $\pi$.

Typical snapshots of pl-systems where chains are colored according to their length are reported in Fig.~\ref{fig:pl-clength}a. In the following, we characterize the emerging chains at different $\epsilon$- and $\Delta$-values according to their typical length, their packing abilities, their nematic order parameter, their characteristic bond angle distribution and their end-to-end distance (see Fig.~\ref{fig:pl-clength} and~\ref{fig:pl_bonds}, and Fig.~\ref{fig:pl_nematic} and Fig.~\ref{fig:pchains} in Appendix~\ref{appendix:pl}).

\textbf{Chain length}. 
The average chain length, $\langle L \rangle$, is defined as the number of monomers in a chain. In all pl-systems,  $\langle L \rangle$ increases as the interaction strength $\epsilon$ rises (see Fig.~\ref{fig:pl-clength}b):  for all pl-types and $\Delta$-values, the formation of chains longer than three particles occur at $\epsilon = -7.2 k_{B}T$. 
In contrast, at $\epsilon = -8.2 k_{B}T$, pl-systems become distinguishable: chains of pl-as tend to be significantly longer than the other types with $\langle L \rangle = 16.92\pm 0.75$ at $\Delta=0.8$, in contrast to $\langle L \rangle = 10.53\pm 0.39$ for pl-center and $\langle L \rangle = 7.02 \pm 0.21$ for pl-s at ${\Delta=0.8}$  Additionally, pl-as chains with $\Delta<0.5$ are notably shorter than chains with $\Delta>0.5$. It is worth noting that the chain length in pl-systems appears to be distributed exponentially (see the inset of Fig.~\ref{fig:pl-clength}b) and thus the error bar was calculated using the standard error of the mean.

\textbf{Packing}.
For all pl-systems, the average packing fraction $\phi=N/(L_x L_y) l^2 \sin(\alpha)$ grows on increasing $\epsilon$ (see Fig.~\ref{fig:pl-clength}c) and it is relatively symmetric with respect to $\Delta=0.5$ (corresponding to pl-center). For $\epsilon\leq -7.2 k_{B}T$, we find that $\phi$ does not depend on the chain type nor on the patch positions, but it is determined only by $\epsilon$. In contrast, at $\epsilon = -8.2 k_{B}T$, when particles start to form longer and longer chains, $\phi$ increases significantly and it becomes dependent on both the chain type and $\Delta$.  At this high energy level, the best packing is achieved by the linear chains of pl-center with $\phi = 0.61\pm 0.02$, as they are able to align seamlessly. As patches move off-center, the staggered chains of pl-s are harder to fit together with respect to the linear chains, resulting in a lower packing fraction that reaches $\phi = 0.49\pm 0.01$ at $\Delta = 0.2$.  The jagged chains in pl-as align better than the staggered chains, with a packing fraction that is still -- but less dramatically -- dependent on $\Delta$ with $\phi = 0.57\pm0.01$ at $\Delta=0.2$. We observe that the packing of jagged chains with small angle bonds is comparable to the packing of jagged chains with big angle bonds.

\textbf{Nematic order parameter}.
While for $\epsilon \leq -7.2k_{B}T$ (corresponding to intermediate packing) the orientations of neighboring chains are independent of each other, at $\epsilon = -8.2k_{B}T$, chains in all pl-systems tend to align with their neighboring ones. We calculate the nematic order parameter in the largest cluster, $S_{\text{largest}}$, (see Appendix~\ref{appendix:pl}) and conclude that all pl-systems pack as nematic fluids, as the fraction of chains in the largest cluster, $f_{\text{largest}}$, typically lies above $0.4$ and the nematic order of these largest clusters is $S_{\text{largest}}\approx 0.6 - 0.85$ (see Fig.~\ref{fig:pl_nematic} in Appendix~\ref{appendix:pl}). 

\textbf{Bond angle distribution}. 
Another way to characterize the emerging chains is the bond angle distribution between neighboring chain elements (see Fig.~\ref{fig:pl_bonds}a)~\cite{Rao2012}. The bonding scenarios can be classified according to the sequence of bond-types along a subchain of three particles. As linear (pl-center) and jagged (pl-as) chains are connected with only one type of bond (on-edge and off-edge p-bonds, respectively), the characteristic bond angles of these systems are close to $0\degree$. The corresponding bonding scenarios are p-p-on (two on-edge p-bonds) for linear chains and p-p-off (two off-edge p-bonds) for jagged chains, where p-p-off bond angles have a slightly broader distribution (green histogram in Fig.~\ref{fig:pl_bonds}a) with respect to p-p-on (blue histogram in Fig.~\ref{fig:pl_bonds}a).  In contrast, bond angles of staggered chains (pl-s) are distributed around three characteristic angles corresponding to three possible bonding scenarios:  p-p-on, p-off$\&$p-on (one off-edge and one on-edge bond) and p-p-off. While the characteristic p-p-on bond angle distribution is independent of $\Delta$ and always peaked around $0\degree$, the distributions of p-off$\&$p-on and p-p-off bond angles are dependent on $\Delta$ (see Fig.~\ref{fig:pl_bond_angles} in Appendix~\ref{appendix:pl}). At $\Delta=0.8$ (reported in red in Fig.~\ref{fig:pl_bonds}a), p-off$\&$p-on is peaked around $30\degree$ and p-p-off is peaked around $55\degree$.
It is important to stress that both the bond angle distributions for off-center systems are significantly wider than for pl-center. This effect is a direct result of the higher bond flexibility of off-center bonds, which we find to rise monotonically the more off-center $\Delta$ becomes. We further study the bond-flexibility by calculating the bonding entropy for all pair configurations as well as the average bend of a chain and its bend range. We define the bonding entropy as the volume of states of a bonded configuration calculated by MC-simulation of two particles in the NVT ensemble (see Appendix~\ref{appendix:entropy}); the average bend of a chain is defined as the mean of the difference in orientation between neighboring chain elements, while the bend range is defined as the standard deviation of the average bend (see Appendix~\ref{appendix:pl}). In general, the more-off center $\Delta$ is, the higher the bonding entropy, the average bend and the bend range. For further details and results see Fig.~\ref{fig:nstates}, \ref{fig:volume_states} in Appendix~\ref{appendix:entropy} and Fig.~\ref{fig:pchains} in Appendix~\ref{appendix:pl}.

\begin{figure*}
\begin{center}
\includegraphics[width=\textwidth]{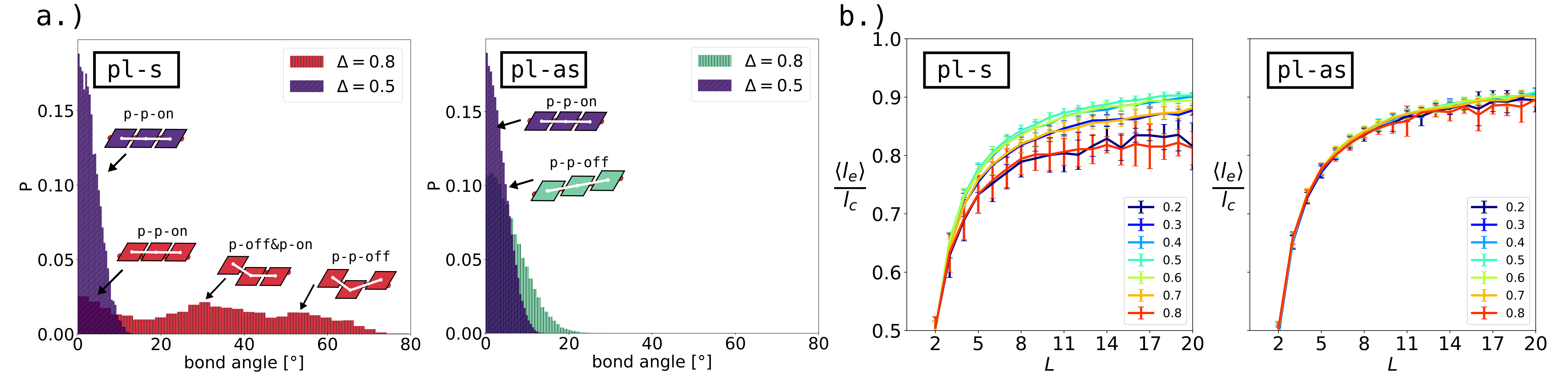}
\caption{\textbf{a.)} Distribution of bond angles for pl-center (blue) pl-s (red) and pl-as (green). Sketches of bonding configurations corresponding to the peaks are added in corresponding colors. The naming scheme of such a configuration takes into account if bonds are parallel or non-parallel as well as if bonds are on- or off-edge, namely ``p-p'': two-parallel bonds; ``p-np'': one parallel and one non-parallel bond; ``np-np'': two non-parallel bonds; ``on'': on-edge bonds, ``off'' off-edge bonds. \textbf{b.)}. The average ratio of the end-to-end distance $l_{e}$ and the contour lengths $l_{c}$, $\langle l_{e}/l_{c} \rangle$, as function of the chain length $L$ and for different patch positions $\Delta$.}
\label{fig:pl_bonds}
\end{center} 
\end{figure*}

\textbf{End-to-end distance}.
The flexibility of a whole chain can be measured by comparing the end-to-end distance $l_{e}$, calculated as the distance between the centers of the first and the last particle in the chain, and the contour length $l_{c}$, which is the length of the chain when maximally stretched. Note that the maximum stretch is reached when mutual rhombi edges are parallel and the patches are maximum distance $2r_{p}$. The contour length $l_{c}$ is then given as the distance vector of the sum of all bond vectors corresponding to this maximum stretch and can be calculated analytically. The bond vectors, referred to as $\mathbf{v}_{max}$ in the following, are dependent on the bond configuration and on $\Delta$. For on-edge bonds (p-on) $\mathbf{v}_{\text{max,on}}$ is constant across all values of $\Delta$ with 
\begin{equation}
\mathbf{v}_{\text{max,on}} = 
\begin{pmatrix}
 l + 2r_{p} \\
 0
\end{pmatrix}.
\label{eq:vmaxo}
\end{equation}
For off-edge bonds (p-off) $\mathbf{v}_{\text{max}}$ depends on $\Delta$ and on whether the bond is closer to the small internal rhombi angles (p-off-s) or closer to the big angles (p-off-b). For p-off-s $\mathbf{v}_{\text{max,s}}$ is given as 
\begin{equation}
\mathbf{v}_{\text{max,s}}(\Delta)  = 
\begin{pmatrix}
l+2r_{p}+\cos(\alpha)|l-2\Delta| \\ 
\sin(\alpha)|l-2\Delta|
\end{pmatrix},
\label{eq:vmaxs}
\end{equation}
while for p-off$_{b}$ we get
\begin{equation}
\mathbf{v}_{\text{max,b}}(\Delta)  = 
\begin{pmatrix}
l+2r_{p}-2\cos(\alpha)|l-2\Delta| \\ 
-\sin(\alpha)|l-2\Delta|
\end{pmatrix}.
\label{eq:vmaxb}
\end{equation}
It is important to note that in pl-center systems, rhombi only bond with $\mathbf{v}_{\text{max,on}}$, while in in pl-as systems, rhombi always bond with the same $\Delta$-dependent bond vector, either $\mathbf{v}_{\text{max,s}}$ (when $\Delta<0.5$) or $\mathbf{v}_{\text{max,b}}$ (when $\Delta>0.5$). In contrast, in pl-s systems all three bond types (p-on, p-off-s, p-off-b) can occur within the same system. Hence, the general expression for calculating the contour length $l_{c}$ for a particular chain is the following
\begin{equation}
\begin{split}
l_{c} & = N_{\text{on}} ||\mathbf{v}_{\text{max,on}}(\Delta)|| \\
& + N_{\text{off}-{s}} ||\mathbf{v}_{\text{max,s}}(\Delta)||\\
& + N_{\text{off}-{b}} ||\mathbf{v}_{\text{max,b}}(\Delta)||,
\end{split}
\end{equation}
where $N_{\text{on}}$ denotes the number of p-on bonds, $N_{\text{off}-{s}}$ is the number of p-off-s bonds and $N_{\text{off}-{b}}$ the number of p-off-b bonds.

The average fraction $\langle l_{e}/l_{c}\rangle$ as a function of the chain length $L$ is a measure of the chain flexibility, where $\langle l_{e}/l_{c}\rangle=1$ is the rigid chain limit. For all pl-systems we observe a monotonous increase of $\langle l_{e}/l_{c}\rangle$ with L, $i.e.$, the longer the chains, the less flexible they are (see Fig.~\ref{fig:pl_bonds}b). In pl-center and pl-as systems, all $\langle l_{e}/l_{c}\rangle$-curves lie on the top of each other for all $\Delta$-values, meaning that these systems possess a comparable, $\Delta$-independent flexibility. In contrast, for pl-s systems $\langle l_{e}/l_{c}\rangle$-curves are dependent on $\Delta$: as soon as $\Delta$ departs from the central value, we observe lower $\langle l_{e}/l_{c}\rangle$-values at fixed $L$  in the whole $L$-range, meaning that the more off-center $\Delta$ is, the more flexible the chains are.

\subsection{Cluster types in ma/mo-systems}\label{sec:mamo}

\begin{figure*}
\centering
\includegraphics[width=0.9\textwidth]{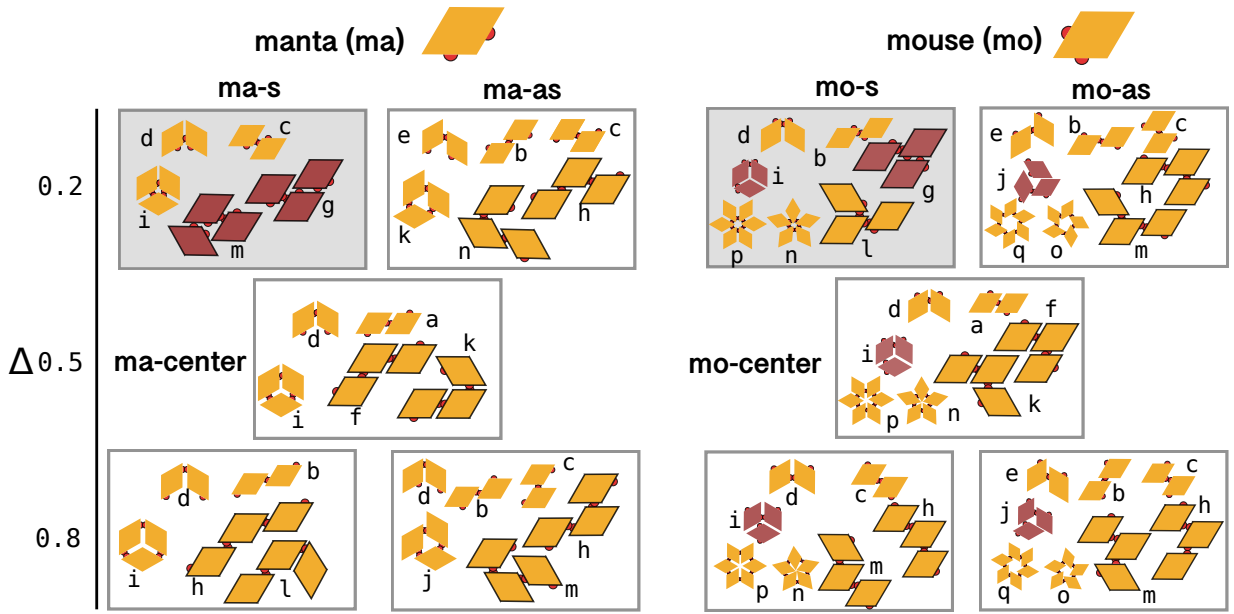}
\caption{Small cluster analysis for manta (ma) and mouse (mo) systems. For each system type (ma on the left, mo on the right, as labeled) there are two columns, one for symmetric (s) topologies and one for asymmetric (as) topologies, as labeled. The rows correspond to different $\Delta$-values: $\Delta = 0.2$ (top), 0.5 (center) and 0.8 (bottom), as labeled. Note the at $\Delta = 0.5$, s-topologies and as-topologies collapse to the center-topology (all patches are on the edge center). Clusters with fully bonded rhombi are colored in orange, while clusters with non-satisfied bonds are colored in burgundy. Systems that do not yield chains are greyed out. The labeling of the different clusters is summarized in the glossary of Fig.~\ref{fig:pl_clusters}. Note that, while we report all possible dimers, the depicted trimers are just examples of how the available bonding patterns at the two-particle level can be combined; (open) boxes and stars are reported as they are the only possible micelles, $i.e.$, minimal loops, that form in ma- and mo-systems, respectively.
}
\label{fig:mamo_clusters}
\end{figure*}

Visual inspection of simulation snapshots shows that in ma- and mo-systems not only chains, but also loops and micelles emerge. In ma-systems, micelles (or minimal loops) consist of three np-bonded particles (referred to as boxes), while in mo-systems they consist of five or six np-bonded  particles (referred to as 5- or 6-stars, respectively). The relative abundance of chains, loops and micelles depends on $\Delta$ and $\epsilon$, but the small cluster analysis reported in Fig.~\ref{fig:mamo_clusters} allows us to already discern which clusters are allowed, $i.e.$, fulfill the given bonding constraints. These constraints follow from the patch configuration (ma- $vs$ mo-systems), the patch topology (s- $vs$ as-topology) and $\Delta$.

In general, s-topologies with $\Delta<0.5$ do not allow for p-bonds in clusters bigger than two, because of bonding incompatibilities (see the relative -- top-left -- panels in Fig.~\ref{fig:mamo_clusters}). Therefore, in both ma- and mo-systems neither chains nor loops are possible when $\Delta < 0.5$, rendering micelles the only self-assembly product. When $\Delta > 0.5$, chains and loops can form because the patch positioning allows for both p- and np--bonds in clusters bigger than two  (see the relative -- bottom-left -- panels in Fig.~\ref{fig:mamo_clusters}). In contrast, we find no such bonding restrictions for as-topologies, and hence chains, loops and micelles are allowed at all $\Delta$-values  (see the relative -- top/bottom-right -- panels in Fig.~\ref{fig:mamo_clusters}). For additional insight into the variety of the emerging clusters, the naming of characteristic dimers and trimers is reported in Fig.~\ref{fig:pl_clusters}. Note that due to the np-off-edge bonding of dimers, micelles in ma-as and mo-as have a hole in the center (for both  $\Delta < 0.5$ and  $\Delta > 0.5$); we refer to the resulting clusters as open-boxes and open-(5- or 6-)stars. 

\begin{figure*}
\begin{center}
\includegraphics[width=0.9\textwidth]{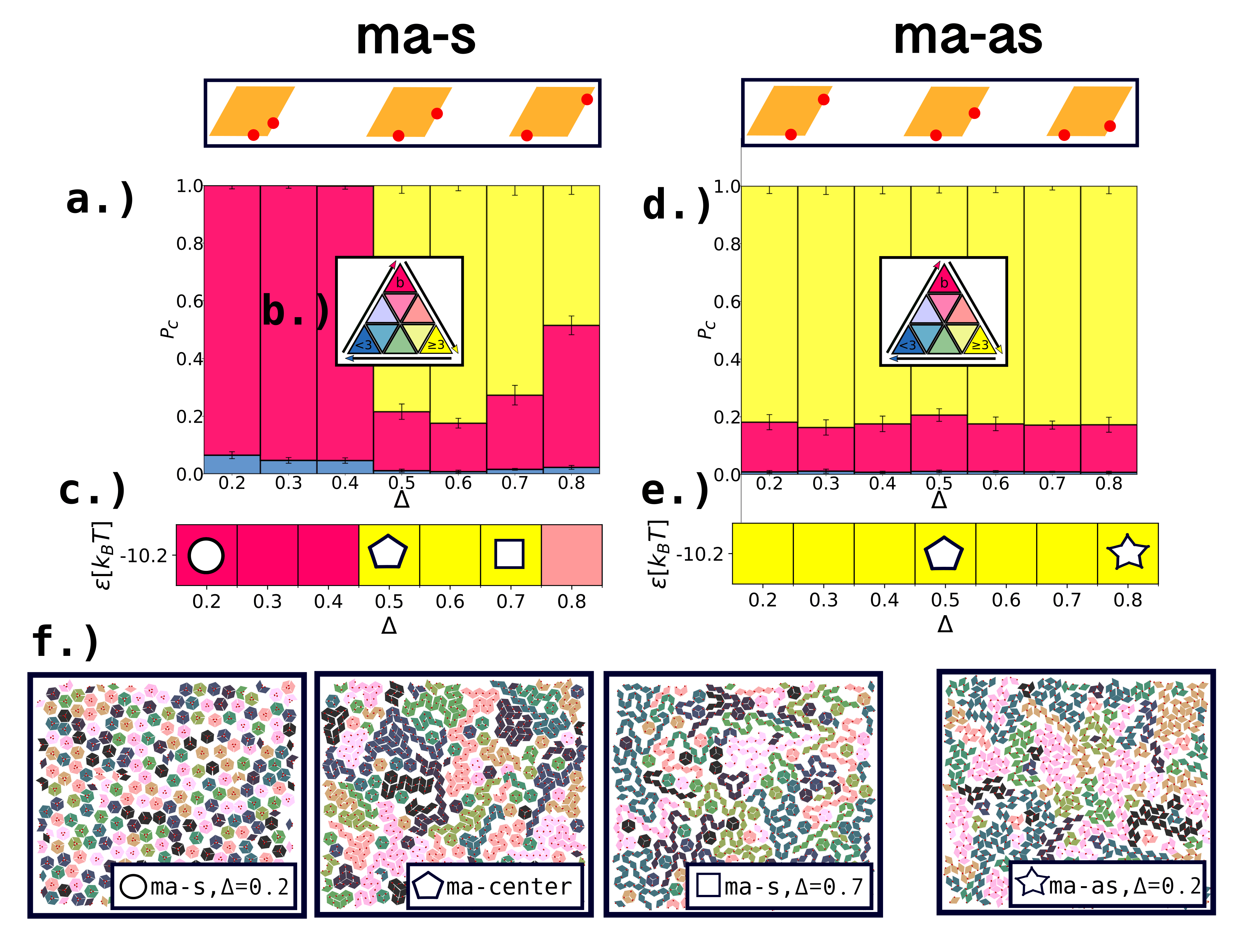}
\caption{Self-assembly products of ma-systems. Top: sketches of ma-s-systems (left panel) and ma-as-systems (right panel) with (from left to right in both panels) $\Delta=0.2, 0.5$ and $0.8$. \textbf{a.)} Histogram of yields for ma-s systems at $\epsilon = -10.2 k_{B}T$ for clusters types of interest: clusters with size $L<3$ (liquid, blue),  boxes (micelles, pink) and non-fully-bonded clusters with size $L\geq 3$ (chains/loops, yellow). \textbf{b.)} The barycentric color triangle maps the yields for these cluster types to the heatmap in \textbf{c.)}. \textbf{d.)} Snapshots of ma-systems (from left to right): ma-s$_{\Delta=0.2}$, ma-center,  ma-s$_{\Delta=0.7}$, ma-as$_{\Delta=0.2}$; the color of the clusters is a guide to the eye to distinguish between different close-by aggregates. \textbf{e.)} Histogram of yields for ma-as systems at $\epsilon = -10.2 k_{B}T$ for clusters types of interest: clusters with sizes $L<3$ (liquid, blue), open-boxes (micelles, purple), and non-fully-bonded clusters with $L\geq 3$ (chains/loops, yellow). The same barycentric color triangle as in \textbf{b.)} maps the yields for these cluster types to the heatmap in \textbf{f.)}.}
\label{fig:ma_chains}
\end{center} 
\end{figure*}

As we are interested in the assembly of chain-like aggregates, we search for the conditions that favor the  predominance of chains over loops and micelles. To do so, we calculate yields of cluster types. In ma-systems, we classify clusters of size $L<3$ as liquid (l), boxes as micelles (m) and non-fully-bonded clusters of size $L\geq 3$ as chains/loops (c). In mo-systems  for $\Delta=0.2$ clusters with $L<5$ are liquid, 5-stars/5-open-stars are micelles and non-fully-bonded clusters of size $L\geq 5$ are chains/loops. For all other $\Delta$ in mo-systems, we selected 6-stars/6-open-stars as micelles (based on their relative abundance with respect to the 5-stars/5-open-stars, see later), and subsequently we define as liquid those clusters with $L<6$ and as chains/loops those non-fully-bonded clusters with $L\geq6$. The yield of a cluster type is defined as the percentage of particles that are part of clusters belonging to the selected cluster type. The obtained yields ($p_l, p_{m}, p_{c}$) are summarized in Fig.~\ref{fig:ma_chains}a for ma-systems and in Fig.~\ref{fig:mo_chains}a for mo-systems. Through a mapping to a barycentric coordinate system (see Appendix~\ref{appendix:mamo}), we obtain the dominant cluster type at each $\Delta$. The resulting heatmaps are displayed in Fig.~\ref{fig:ma_chains}c/g for ma-systems and in Fig.~\ref{fig:mo_chains}d/h for mo-systems. Note that in Appendix~\ref{appendix:mamo}and Fig.~\ref{fig:loops} we discuss and show histograms at selected $\Delta$-values where chains and loops are distinguished.

From Fig.~\ref{fig:ma_chains}a and c, we observe that in ma-s-systems, boxes are the dominant cluster type for $\Delta<0.5$ with yields above $0.9$, as bonding constraints do not allow chains. For ma-center, the box yield drops to $0.204\pm 0.026$, while the chain yield is $0.746\pm 0.0234$ and the loop yield is $0.049\pm 0.011$ (see Fig.~\ref{fig:loops} in Appendix~\ref{appendix:mamo} for yield distributions where chains and loops are distinguished). Moving to higher $\Delta$, we observe that the box yield is minimal at $\Delta=0.5$ and $\Delta=0.6$ and then rises again to reach $0.492\pm 0.032$ at $\Delta=0.8$. In ma-as systems (see Fig.~\ref{fig:ma_chains}f and g), we find chains to be the dominant cluster type for all $\Delta$-values.  Furthermore, we observe the yields to be independent of $\Delta$ for all cluster types, with a chain yield of $\approx 0.75$, a open-box yield of $\approx 0.2$ and a loop yield of $\approx 0.05$ (see again Fig~\ref{fig:loops} in Appendix~\ref{appendix:mamo}).

In mo-s systems (reported in Fig.~\ref{fig:mo_chains}a and d), stars are the prevalent cluster type for $\Delta<0.5$ where 5-stars are dominant at $\Delta=0.2$ with a yield $0.837 \pm 0.041$ and 6-stars are dominant at $\Delta=0.3$ and $0.4$ with a yield of $0.580 \pm 0.031$ and $0.554 \pm 0.044$, respectively. For mo-center, chains are dominant with a yield of $ 0.976$, while the 6-star yield is only $0.024$ and loops are not present (Fig.~\ref{fig:loops} in Appendix~\ref{appendix:mamo}). For $\Delta>0.5$ mo-s chains remain the most prevalent cluster type with a yield over $0.95$, while the star yield remains below $0.05$. In contrast to mo-center, when $\Delta >0.5$ loops are present but with yields of below $0.05$ (Fig.~\ref{fig:loops} in Appendix~\ref{appendix:mamo}). In mo-as systems (reported in Fig.~\ref{fig:mo_chains}g and h), chains are the dominant cluster types at all $\Delta$ with yields above $0.95$ and independent of $\Delta$. Open-stars and loops only reach yields below $\approx 0.05$.

We conclude that for both ma- and mo-systems, the s-topology favors micelles when $\Delta<0.5$ and chains/loops for $\Delta \ge 0.5$, while the as-topology favors chains/loops over the whole $\Delta$-range. Moreover, when micelles are disfavored, chains prevail over loops. We now focus on the regimes were chains are dominant and characterize the emerging aggregates, as the snapshots reported in Fig.~\ref{fig:ma_chains}d and Fig.~\ref{fig:mo_chains}e suggest chains in different systems have different features. 

\begin{figure*}
\begin{center}
 \includegraphics[width=0.9\textwidth]{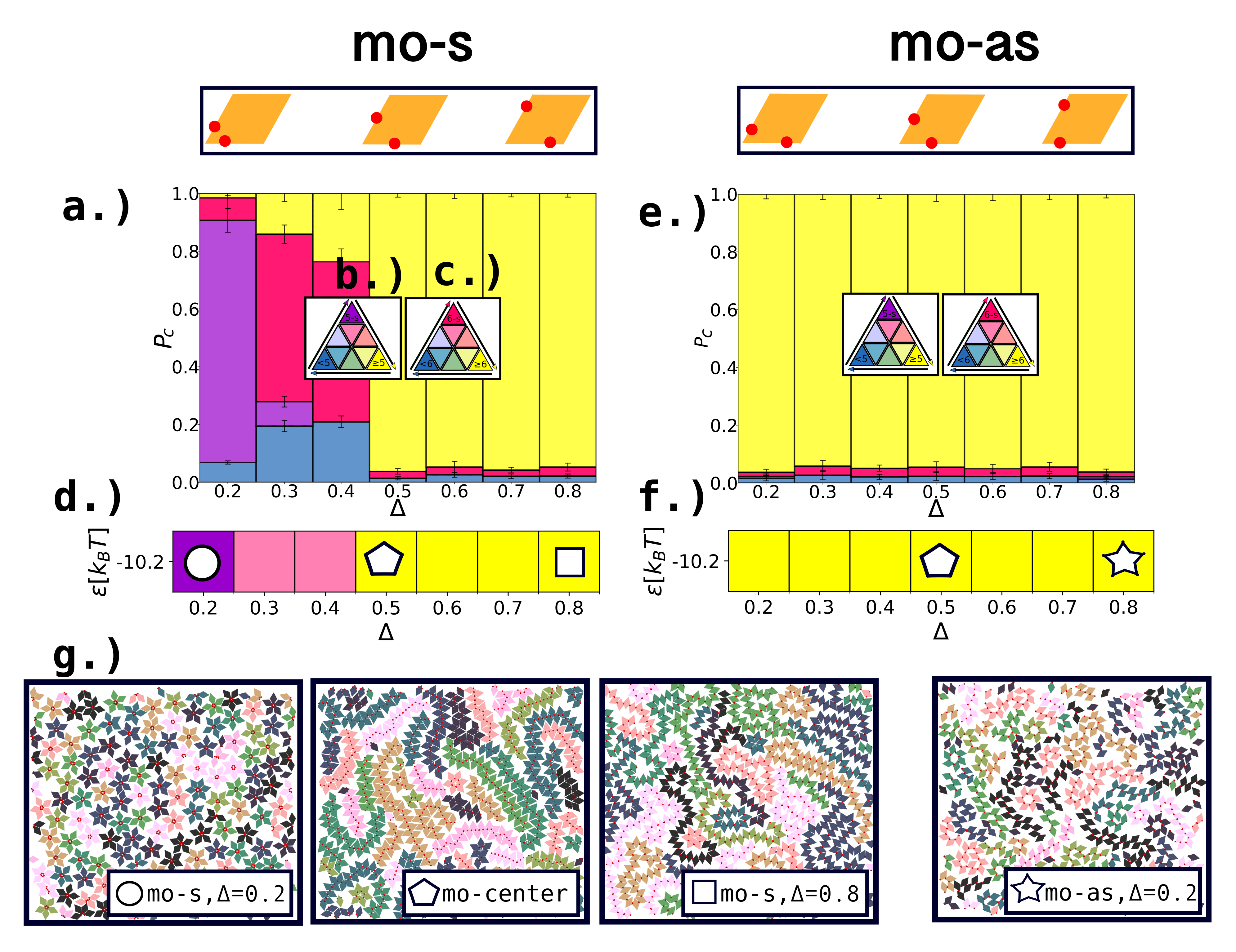}
  \caption{Self-assembly products of mo-systems. Top: sketches of mo-s-systems (left panel) and mo-as-systems (right panel) with (from left to right in both panels) $\Delta=0.2, 0.5$ and $0.8$. \textbf{a.)} Histogram of yields for mo-s systems at $\epsilon = -10.2 k_{B}T$ for clusters types of interest: clusters with sizes $L<5$ (liquid,blue), 5-stars (micelles of five particles, purple), 6-stars (micelles of six particles, pink) and clusters with sizes $L\geq 5$ (chains/loops, yellow). \textbf{b.)} The barycentric color triangle maps the yields for the cluster types liquid (blue), 5-stars (purple) and chains/loops (yellow) for the patch position $\Delta=0.2$ to the heatmap in \textbf{d.)}. \textbf{c.)} The barycentric color triangle maps the yields for the cluster types liquid (blue), 6-stars (pink) and chains/loops (yellow) for $\Delta>0.2$ to the heatmap in \textbf{d.)}. \textbf{e.)} Snapshots of mo-systems (from left to right): mo-s$_{\Delta=0.2}$, mo-center,  mo-s$_{\Delta=0.8}$, mo-as$_{\Delta=0.2}$; the color of the clusters is a guide to the eye to distinguish between different close-by aggregates. \textbf{f.)} Histogram of yields for mo-as systems at $\epsilon = -10.2 k_{B}T$ for clusters types of interest: clusters with sizes $L<5$ (liquid, blue), 5-open-stars (micelles of five particles, purple), 6-open-stars (micelles of six particles, pink) and clusters with $L\geq 5$ (chains/loops, yellow). The barycentric color triangle in \textbf{b.)} maps the yields for the cluster types liquid (blue), 5-open-stars (pink) and chains/loops (yellow) for the patch position $\Delta=0.2$ to the heatmap in \textbf{g.)} The barycentric color triangle in \textbf{c.)} maps the yields for the cluster types liquid (blue), 6-open-stars (pink) and chains/loops (yellow) for $\Delta > 0.2$ to the heatmap in \textbf{g.)}.}
\label{fig:mo_chains}
\end{center} 
\end{figure*}

\subsection{Chains in ma/mo systems}

The emerging chains in ma- and mo-systems differ from each other with respect to their bonding pattern and therefore also with respect to their appearance and physical properties. 
From the visual analysis of simulation snapshots (reported in Fig.~\ref{fig:ma_chains}d and Fig.~\ref{fig:mo_chains}e), we can already infer some characteristic features which distinguish the chains emerging in the different systems. In ma-center and ma-s (with $\Delta>0.5$), the constraints imposed by on-edge p-bonds and off-edge np-bonds (see Fig.~\ref{fig:mamo_clusters}) are such that the rhombi tend to turn their vertices with the large angles outward, giving ma-center and ma-s chains a pipe-like appearance. In ma-as on the other hand, rhombi bind exclusively off-edge (for both p- and np-bonds), resulting in jagged chains, where both, small and large angled vertices stick out. In mo-center and mo-s, the constraints imposed by on-edge p-bonds and off-edge np-bonds (see again Fig.~\ref{fig:mamo_clusters})  are such that the rhombi turn their vertices with small angles outward, making the chains spike-like in appearance. In mo-as, again p- and np-bonds  are both off-edge, thus leading to jagged chains. 

Similar to the analysis of pl-systems, also for ma- and mo-systems we characterize the emerging chains at different $\Delta$-values (note that for ma- and mo-systems $\epsilon=-10.2 k_BT$) according to their typical length, their packing abilities, and their characteristic bond angle distribution (see Fig.s~\ref{fig:mamo_clength} and~\ref{fig:mamo_bond_angle}, and Fig.~\ref{fig:moma_flexibility} in Appendix~\ref{appendix:mamo}. Note that we do not show the nematic order parameter analysis nor the end-to-end distance as the kinks that characterize the emerging chains do not allow for a meaningful definition of the chain backbone.  

\textbf{Chain length.}
In both ma- and mo-systems chain lengths appear to be distributed exponentially (see Fig.~\ref{fig:mamo_clength}b). Additionally average chain lengths increase or decrease as a function of $\Delta$, symmetrically with respect to $\Delta=0.5$ (see Fig.~\ref{fig:mamo_clength}a).  In ma-center, chains have an average length of $\langle L \rangle = 12.97\pm 0.46$. In ma-s (with $\Delta>0.5$), the chain length decreases monotonically with $\Delta$ and for $\Delta=0.8$ the average chain length is $\langle L \rangle = 7.37\pm0.26$. In ma-as, on the other hand, the chain length increases monotonically with $\Delta$ and it is $\langle L \rangle = 15.76 \pm 0.69$ for $\Delta=0.8$. In mo-center, chains have an average length of $\langle L \rangle =17.33 \pm 0.64$ and in both mo-s and mo-as chains increase with respect to mo-center, and reach $\langle L \rangle = 21.10\pm 1.10$ at $\Delta=0.8$ for mo-s and  $\langle L \rangle = 23.25 \pm 1.37$ for mo-as.  We can conclude that, in mo-systems chains emerge that are on average longer than those in ma-systems, the longest chains emerging in both cases at extreme $\Delta$-values for the asymmetric topology (jagged chains). On the other hand, the shortest chains are observed for ma-s systems at $\Delta=0.8$ (pipe-like chains).

\textbf{Packing.}
As a general rule, the packing fraction for ma- and mo-center is higher than for their respective off-center topologies (see Fig.~\ref{fig:mamo_clength}c). This is due to  the fact that when patches are placed on the edge center only on-edge bonds can form (either p- or np-ones), which enables the chains to fit into each other easily,  leading to a high packing fraction of $\phi = 0.659 \pm 0.010$ for ma-center and $0.648\pm0.006$ for mo-center. Since the bonding becomes more off-edge as $\Delta$ becomes  more extreme, the packing fraction goes down monotonously from $\Delta=0.5$, at $\Delta=0.8$ $\phi = 0.566 \pm 0.004$ for ma-s, $\phi = 0.575 \pm 0.005$ for ma-as, $\phi=0.576\pm0.004$ for mo-s and $\phi=0.550\pm0.003$ for mo-as. While the best packing is achieved by ma-as-systems over the whole $\Delta$-range, the worst packing is not associated to a specific system.

\begin{figure*}
\begin{center}
\includegraphics[width=\textwidth]{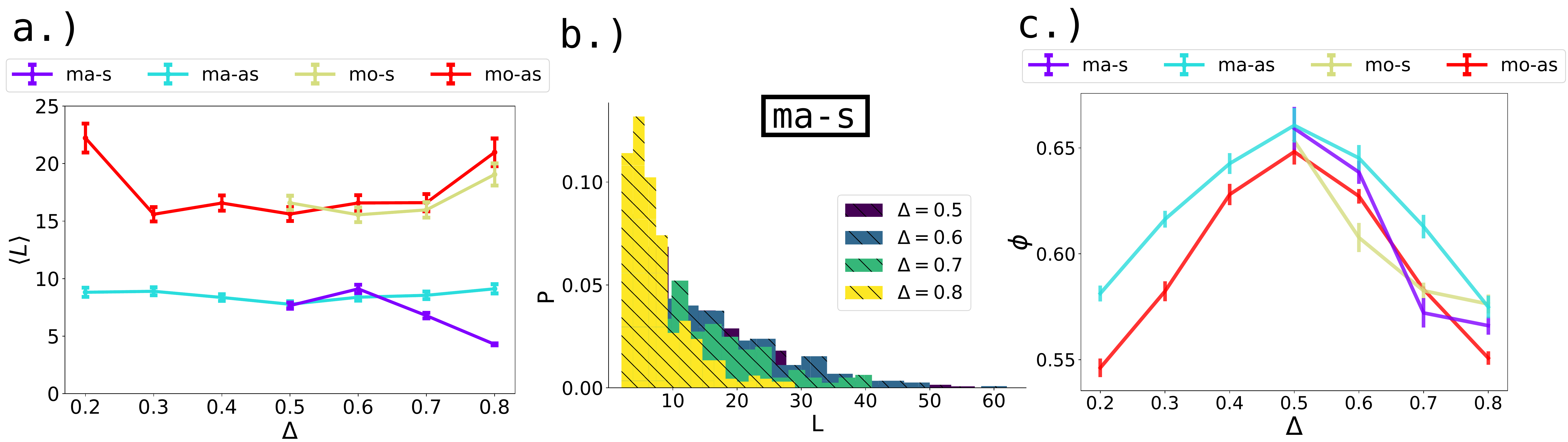}
\caption{\textbf{a.)} $\langle L \rangle$ as a function of $\Delta$ for both ma- and mo-systems, \textbf{b.)} Cluster size distribution of ma-s, \textbf{c.)} $\phi$ as a function of $\Delta$ for both ma- and mo-systems.}
\label{fig:mamo_clength}
\end{center} 
\end{figure*}

\textbf{Bond angle distribution.}
Bond angles are dependent on the possible bonding configurations within one particle class (pl, ma, mo), on the topology (s, as) and on the value of $\Delta$. 
In both ma- and mo-systems, three bonding configurations are possible: p-p (two parallel bonds), p-np (one parallel and one non-parallel bond) and np-np (two non-parallel bonds).
Additionally, depending on the particle class and on $\Delta$, we find on- or off-edge bonds, the last ones of two-types: p-off-s and p-off-b. 

The set of possible bond-angles determines the kinks in the backbone and therefore also system properties, such as packing fraction and average chain length. Each of the investigated systems is characterized by a unique bond angle distribution.  

For ma-center the bond-angle distribution is bimodal with one peak at $60\degree$ (p-p-on bonding) and one at $90\degree$ (p-np-on).  The on-edge bond together with the sharp angles, give ma-center chains the appearance of double stranded chains. 
In ma-s, where we increase $\Delta$ symmetrically off-center, the p-p-off and np-on$\&$p-off peaks are shifted to smaller angles compared to the on-edge bonds of ma-center. Additionally, we observe a third peak, the np-np-on peak at $\approx 110\degree$, which represents boxes with one broken bond. The off-edge bonding gives ma-s chains the appearance of single stranded chains. 
In ma-as, the p-p-off and the np-np-off peak shift and spread into each other to form one diffuse mode. Similar to to ma-s, we find np-np-off configurations, $i.e.$, open-boxes with one broken bond. 

In mo-center, we observe a bond angle distribution with three peaks, the p-p-on at $120\degree$, the p-np-on at $90\degree$ and the np-np at $60\degree$.  The on-edge bonds, combined with the sharp bond-angles gives mo-center a double stranded appearance. 
In mo-s, as $\Delta$ is moved symmetrically off-center, all three peaks merge into one, at $\approx 50\degree$ for $\Delta=0.8$.
In mo-as, as $\Delta$ is moved asymmetrically off-center, we observe that the three peaks extend into each other, forming one diffuse mode at $\Delta=0.2$.

In all systems, the spread of the distributions widens the more off-center $\Delta$ becomes.  This wider spread is directly related to the fact that the average bend and bend range rise as  $\Delta$ moves off-center, which in itself originates from the higher bond entropy of off-center pair configurations (see Fig.~\ref{fig:moma_flexibility} in Appendix~\ref{appendix:mamo} for plots of average bend and bend range in ma/mo systems and Appendix~\ref{appendix:entropy} for a calculation of the bonding entropy).

\begin{figure}
\begin{center}
\includegraphics[width=\columnwidth]{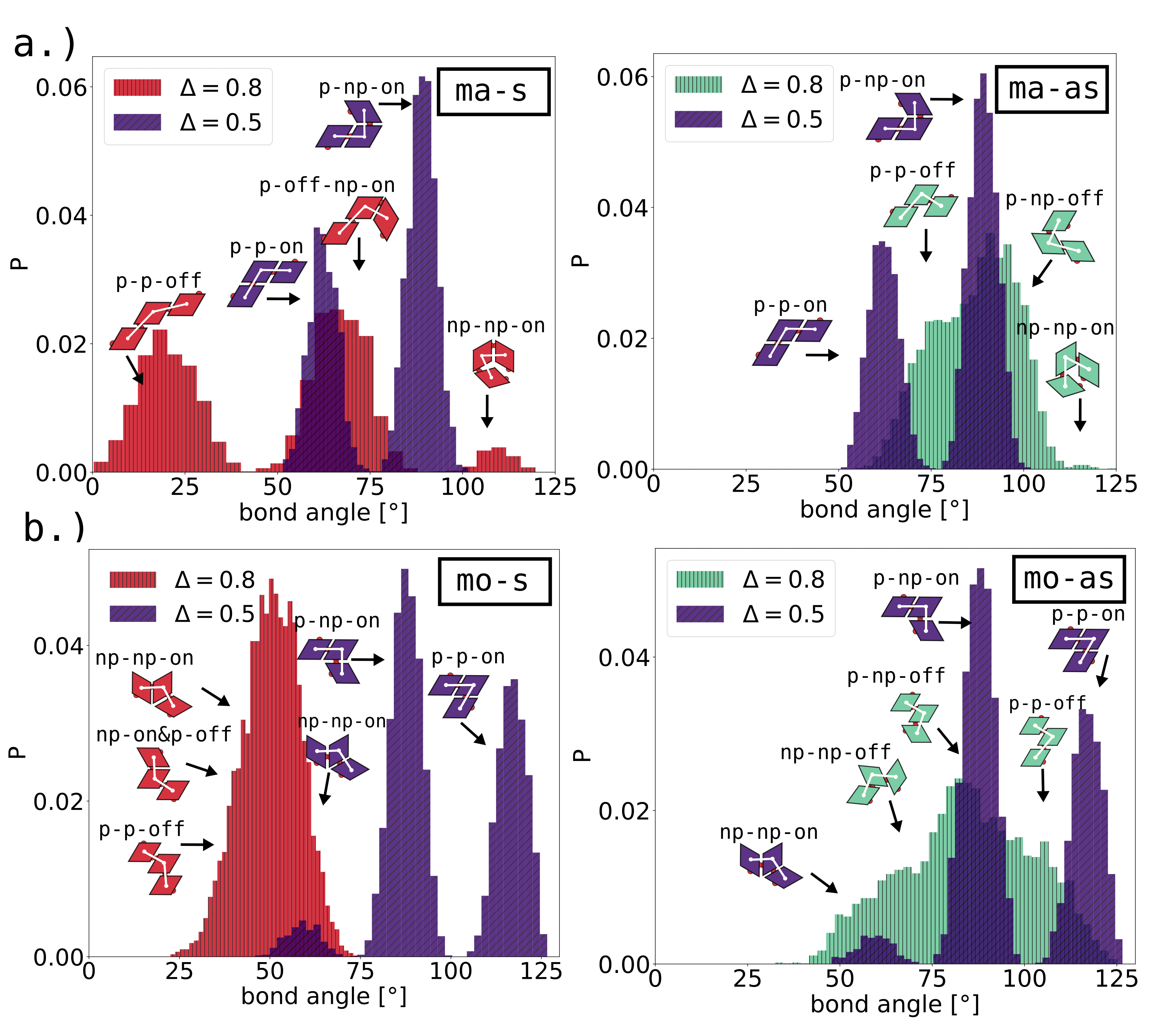}
\caption{\textbf{a.)} Bond angle distributions for ma-center (blue), ma-s (red) and ma-as (green). Sketches of bonding configurations corresponding to the peaks are added in corresponding colors. The naming scheme of such configuration takes into account if bond are parallel or non-parallel as well as if bonds are on- or off-edge, namely ``p-p'': two-parallel bonds; ``p-np'': one parallel and one non-parallel bond; ``np-np'': two non-parallel bonds; ``on'': on-edge bonds, ``off'' off-edge bonds. \textbf{b.)} Bond angle distributions for mo-center (blue), mo-s (red) and mo-as (green). Sketches of bonding configurations are added in corresponding colors. The naming scheme is the same as in panel \textbf{a.)}.
}
\label{fig:mamo_bond_angle}
\end{center} 
\end{figure}

\section{Conclusion}\label{sec:conclusions}

In this paper we explore the formation of chains emerging in systems of rhombic platelets decorated with two mutually-attractive patches on distinguished edges. We consider the effect of the patch interaction strength and of the patch arrangement -- quantified by a patch topology (either symmetric or asymmetric) and a patch position parameter, $\Delta$ -- on the emerging particles assemblies.  For patches placed on opposite edges (referred to as parallel patchy rhombi), chains are the only possible assembly product and they emerge at all the investigated interaction strengths, for both topologies, at all $\Delta$-values.  When patches are placed on adjacent edges (either around the small or the big angle, referred to as manta and mouse patchy rhombi, respectively), chains emerge only at high interaction strengths, where a competition between chains and loops is observed when patches are either in the center of the edges or symmetrically displaced with respect to their common vertex by a $\Delta >0.5$. In contrast, for symmetric patch topologies, chains are only allowed when $\Delta >0.5$, as for $\Delta<0.5$ only loops and micelles ($i.e.$, minimal loops) can form. 

Parallel patchy rhombi form well-defined chains with distinguished properties according to the patch topology: we distinguish between linear, staggered and jagged chains, based on their average length, packing abilities, ordering trends and the flexibility properties. In general, chains of parallel patchy rhombi pack as nematic fluids, with center patch topologies packing better than off-center ones. This is important for a possible second-stage of assembly. Future investigations might focus on the close-packing limit of the different chains and on the percolation properties of these systems, in view of potential applications. 

In contrast, chain-like assembly emerging in manta and mouse systems -- despite being characterized by average lengths and packing fractions comparable to those observed in parallel patchy rhombi systems and by sharp, characteristic bond angle distributions -- cannot be classified as nematic fluids. The flexibility properties of the chains emerging in these systems must be seen in terms of sequences of kinks; depending on the patch positioning the kinks determine the characteristics of the chains. 

Finally, we note that the choice of the number of patches plays an important role in determining the final assembly products. Recently, we have shown that patchy rhombi with four interaction sites are able to grow extended tilings in two-dimensions~\cite{Karner2019}. Through choosing  particular patch topologies, monolayers can be assembled with identical lattice geometry but different porosity, from a close-packed arrangement to an open lattice. Interestingly, the open lattices observed in Ref.~\cite{Karner2019} result from the second-stage assembly -- due to the two additional patches -- of the open micelles (both boxes and stars) observed here. A more thorough analysis of micelles is reported in Ref.~\cite{micelles}.

\section*{Acknowledgements}
CK and EB acknowledge support from the Austrian Science Fund (FWF) under Proj. No. Y-1163-N27. Computation time at the Vienna Scientific Cluster (VSC) is also gratefully acknowledged.

\appendix 

\renewcommand\thefigure{\thesection.\arabic{figure}}    
\setcounter{figure}{0}    

\section{Bonding entropy}
\label{appendix:entropy}

We compare the bonding entropy of bonded pair of particles for different bonding configurations and patch positions $\Delta$.
In general, the bonding entropy is given as the volume of states of a bonding configuration. We estimate the volume of states by conducting a MC-simulation of two particles in the NVT ensemble. 
The particles are initialized in a bonded state and moves that break the bond are rejected. To harvest possible bonded states it suffices to move one of the two particles while the other stays fixed.

\begin{figure}
\begin{center}
\includegraphics[width=1.0\columnwidth]{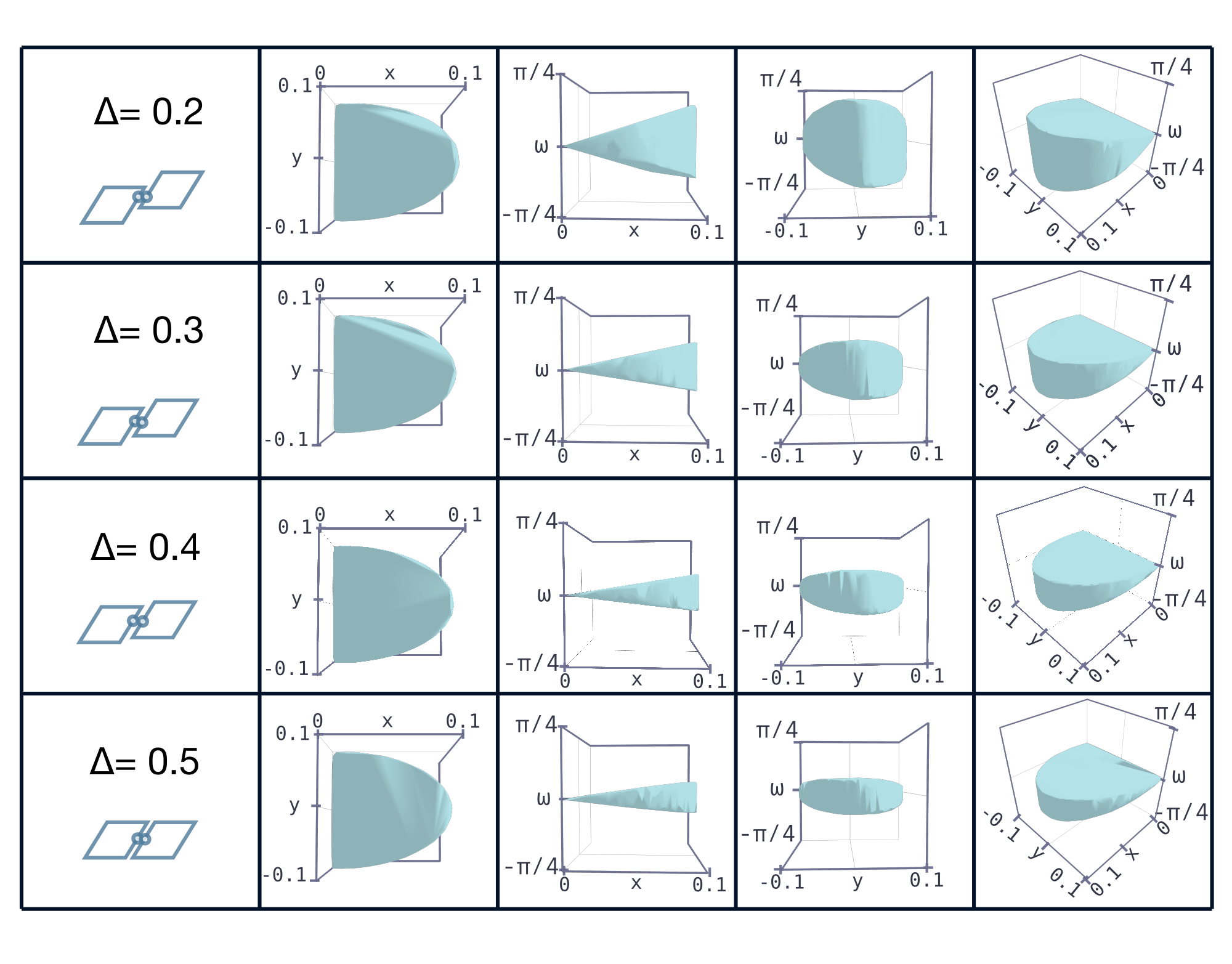}
\caption{The shape of the volume of states for parallel (p) off-edge bonded pair of particles. The state space is given by the patch-patch distance $(x,y)$ and the difference in orientation $\omega$.
The convex shape was calculated with the qhull library from the scatter of numerically obtained bonded states $(x,y,\omega)$. Every row corresponds to a different $\Delta$-value, while every column depicts a particular axis view of the three-dimensional shape (from left to right): $(x,y)$-axis, $(x,\omega)$-axis, $(y,\omega)$-axis, full 3d view.}
\label{fig:volume_states}
\end{center} 
\end{figure}

The bonded states are fully described by the patch-patch distance vector $(x,y)$ and the difference in orientation $\omega$. To compare the bonding entropy of different configurations we calculated the volume of states relative to a model system where the moving particle can rotate freely around a fixed, free-standing patch. We discretize the state space of this free-rotating model system, which is a cuboid with $(x,y,\omega)\in [-0.1,0.1]\times[0,0.1]\times[-\pi,\pi]$, into cubes with a side-length of $0.001$.
Subsequently we evaluate the fraction of discretized cubes occupied by state points and thus estimate the relative volume of states. 
The resulting relative volumes of state for different configurations are given in Fig.~\ref{fig:volume_states}. 


\begin{figure}
\begin{center}
\includegraphics[width=1.0\columnwidth]{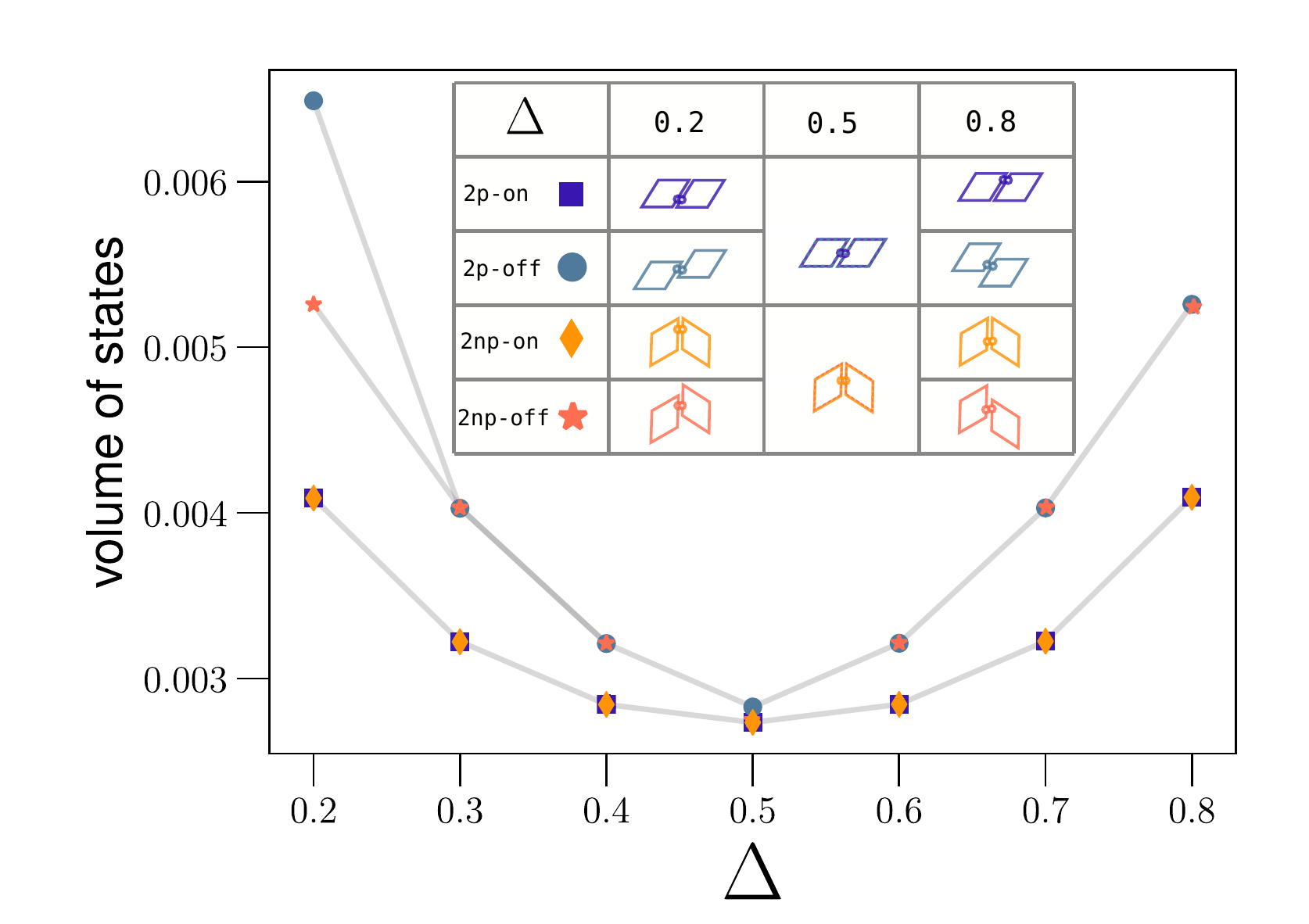}
\caption{The volume of states for different bonding configurations as a function of the patch position $\Delta$. The nomenclature is: 2p-on (parallel (p) on-edge bond, dark blue), 2np-on (non parallel (np) on-edge bond, orange), 2p-off (p-off-edge bond, light blue), 2np-off (np-off-edge bond, red). Inset: sketches of all bonding configurations for different $\Delta$-values. Note that the color scheme is the same as for the volume of states plot.}
\label{fig:nstates}
\end{center} 
\end{figure}


We find that the parallel and non-parallel center positions ($\Delta=0.5$) occupy the least volume of states $0.002734$ for both p-center and np-center (see Fig.~\ref{fig:nstates}).
While the volume of states increases quadratically for all configurations as $\Delta$ moves off-center, 
off-edge bonds yield a higher volume of states than on-edge bonds. 
The volume of states increases for off-center bonds because states with more extreme differences in orientation become available as the patches are moved off-center. 

The volume of states of p- and np-on-edge bonds is almost identical for all $\Delta$-values, while for off-edge bonds it is identical for all $\Delta$-values except for $\Delta=0.2$, where p-bonds yield a much higher volume of states. 

In Fig.~\ref{fig:volume_states} we show the shape of the volume of states for off-edge p-bonds at different $\Delta$-values.
We observe that the overall shape is retained for all $\Delta$-values and the number of states in the $(x,y)$ plane remains constant. However, as $\Delta$ becomes more off-center, more extreme differences in orientation $\omega$ are allowed and this is reflected by a volume growth in the $\omega$ direction.

\section{Pl-systems}
\label{appendix:pl}
\setcounter{figure}{0}    

This section discusses the orientational order, the bond angle distribution and the bond flexibility of pl-chains.

\subsection{Nematic order}

\begin{figure*}
\begin{center}
\includegraphics[width=0.7\textwidth]{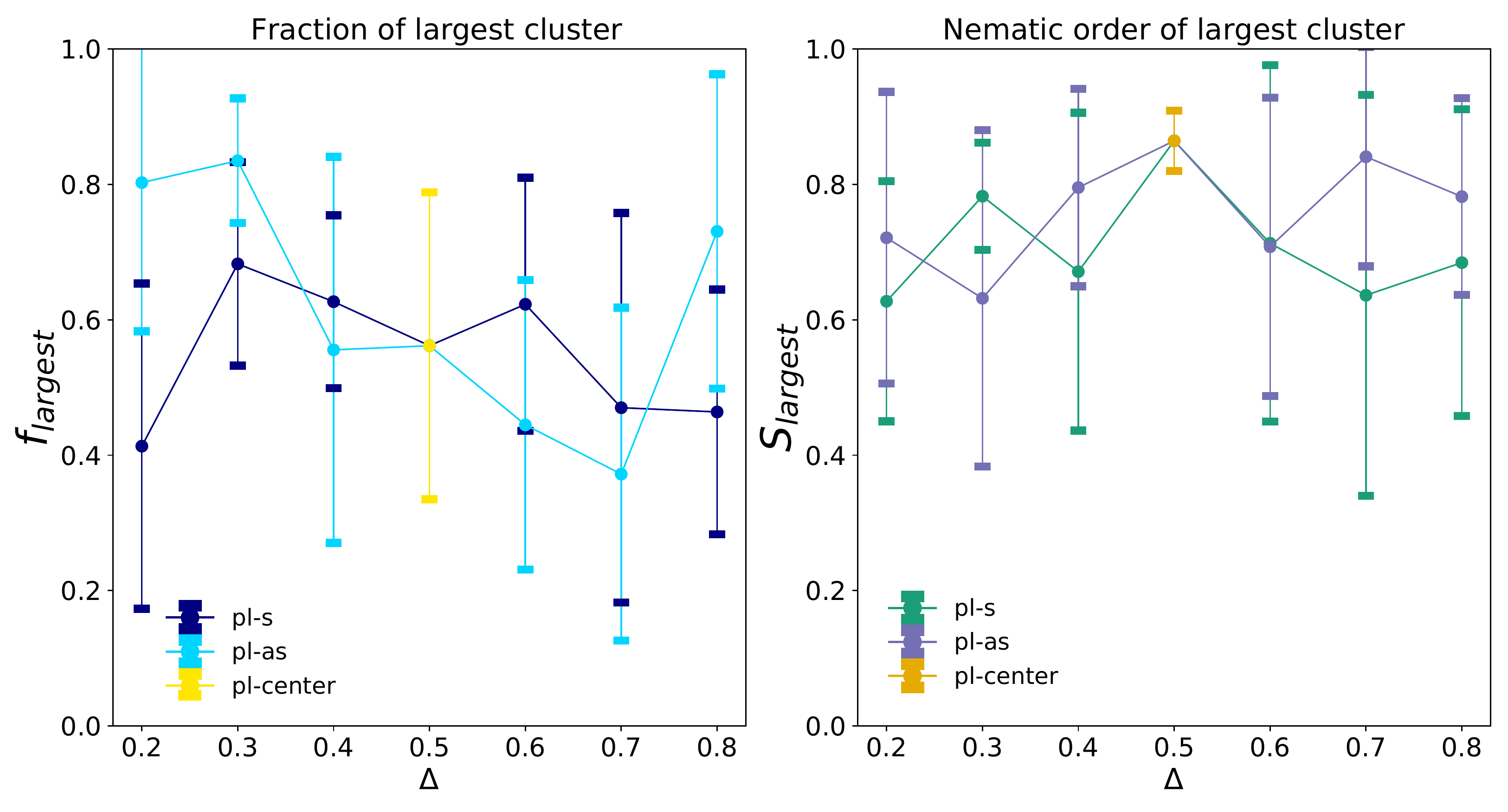}
\caption{Nematic order in pl-systems. Left: The fraction of chains in the largest nematic cluster, $f_{\text{largest}}$,  as a function of $\Delta$ for pl-center (yellow), pl-s (dark blue) and pl-as (light blue). Right: Nematic order parameter of the largest cluster, $S_{\text{largest}}$, as function of $\Delta$ for pl-center (yellow), pl-s (green), pl-as (blue).}
\label{fig:pl_nematic}
\end{center} 
\end{figure*}


At interaction strengths $\epsilon = -8.2 k_{B}T$ pl-chains exhibit a packing fraction $\phi$ between $0.5$ and $0.6$ and the chains align orientationally for all topologies and $\Delta$-values. 
We calculated the fraction of chains in the largest orientationally ordered cluster and the degree of orientational - nematic - ordering within the largest cluster.
First, we define the normalized end-to-end distance vector $\mathbf{u}$ as director or chain orientation vector.
To determine whether a chain $i$ is in an orientationally ordered environment we calculate the local nematic order $S(i)$ according to \cite{Frenkel1985, Cuetos2007} with  
\begin{equation}
S(i) = \frac{1}{N_{b}}\sum_{j=0}^{j=Nb} 2\| \mathbf{u}_{i}\cdot \mathbf{u}_{j}\|^{2} - 1,  
\end{equation}
where $N_{b}$ is the number of neighbouring chains.
We define chains $i$ and $j$ to be neighbours if a monomer in chain $i$ has a center-to-center distance less than $2\cdot l$ to a monomer in $j$. 
A chain is considered nematically ordered if $S(i)>0.4$. 
The criterium that two neighbouring chains are in the same nematic cluster is $\| \mathbf{u}_{i}\cdot \mathbf{u}_{j}\| > 0.85$.
With this criterium we can find all clusters and determine the largest.

The results are summarized in Fig.~\ref{fig:pl_nematic}.
We observe that the fraction of chains in the largest cluster is above $0.4$ and the nematic ordering of the largest cluster is above $0.6$ for all topologies and $\Delta$-values. Hence, we deduce that all pl-chains exhibit nematic ordering at $\epsilon = -8.2 k_{B}T$. However, the large error-bars both on the fraction of particles in the largest cluster and on the nematic order parameter prohibit any further conclusions. 

\subsection{$\Delta$-dependence of bond angle distribution}
\begin{figure}
\begin{center}
\includegraphics[width=0.8\columnwidth]{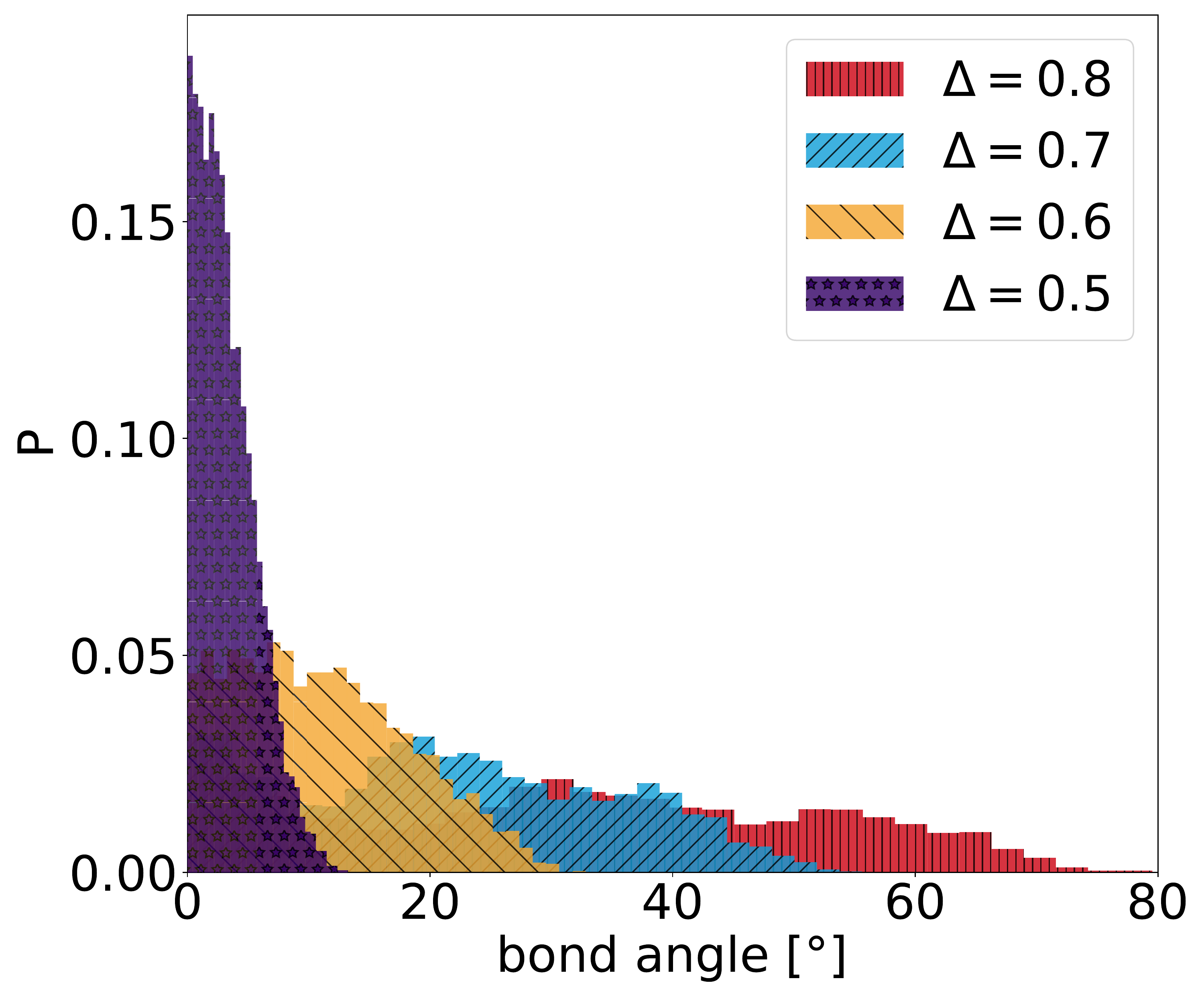}
\caption{Bond angle distributions for pl-s systems, for different values of $\Delta$, as labeled.}
\label{fig:pl_bond_angles}
\end{center} 
\end{figure}

In Fig.~\ref{fig:pl_bond_angles} we plot the bond angle distribution for all $\Delta>0.5$ for pl-s systems to illustrate the $\Delta$-dependence of bond angles. 
See main text for calculation details and discussion. 

\begin{figure*}
\begin{center}
\includegraphics[width=0.8\textwidth]{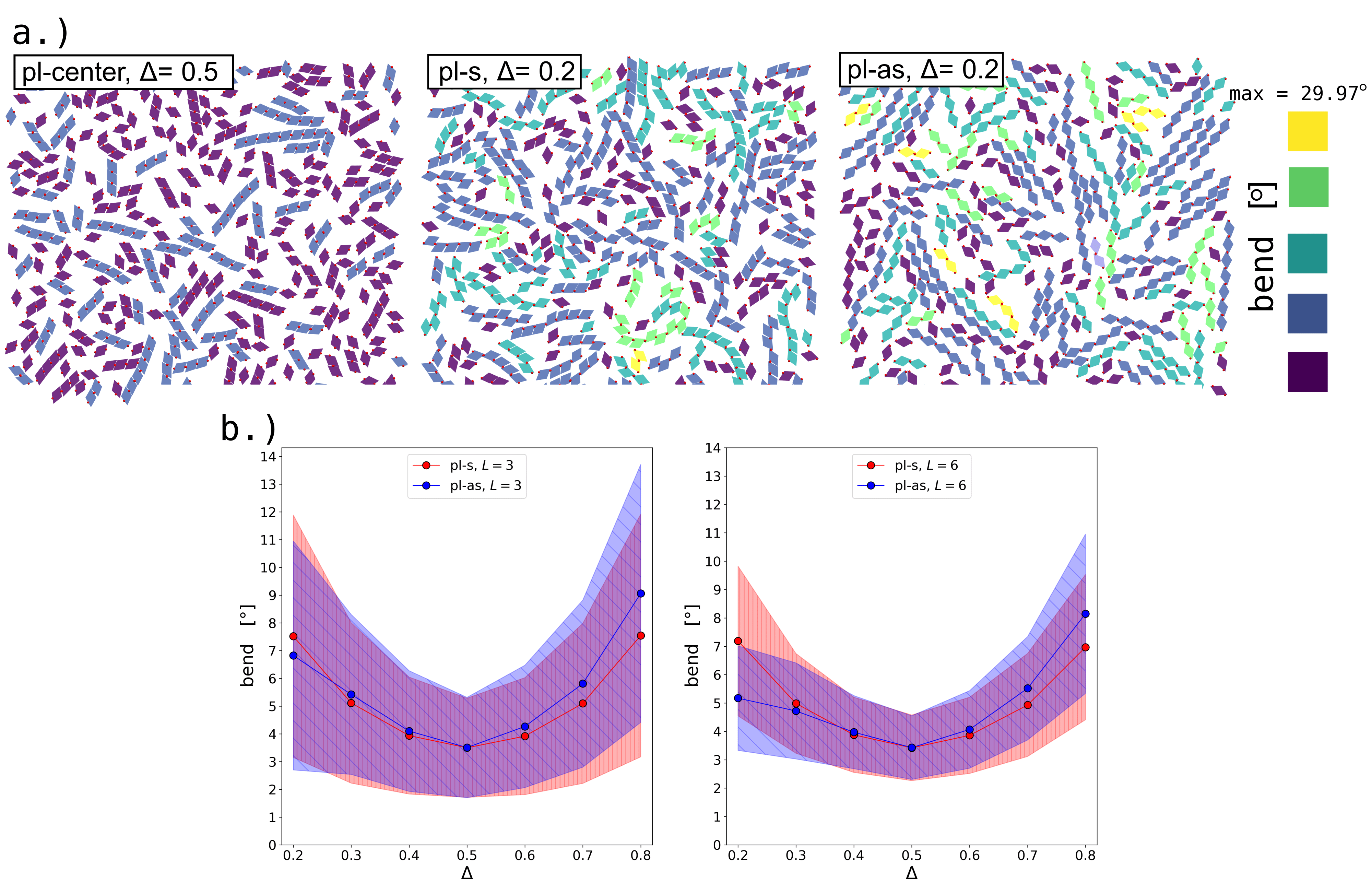}
\caption{\textbf{a.)} Simulation snapshots for pl-center, pl-s with $\Delta=0.2$ and pl-as with $\Delta = 0.2$ at $\epsilon = -7.2 k_{B}T$. (from left to right). Chains are colored according to their bend. The color map is perceptionally uniform and normalized to the maximum bend observed ($i.e.$, $29.972 \degree$). \textbf{b.)} Plot of the average bend and its bend range as function of $\Delta$ for chain lengths $L=3$ (left) and $L=6$ (right) in pl-s and pl-as, both at $\epsilon = -7.2 k_{B}T$.}
\label{fig:pchains}
\end{center} 
\end{figure*}


\subsection{Bond flexibility}
We define the bend of a chain as the mean of the difference in orientation between neighbouring chain elements.
To compare different pl-systems, we evaluated the average bend of chains for every system and as function of $\Delta$. We defined the bend range
as the standard deviation of the average bend (see Fig.~\ref{fig:pchains}b).
In the snapshots of Fig.~\ref{fig:pchains}a chains are colored according to their bend.

In general, the number of available positions and orientations in a bonded state increases the more off-center $\Delta$ becomes (see Fig.~\ref{fig:nstates} and Fig.~\ref{fig:volume_states}).
As a result, pl-center chains are the stiffest, $i.e.$, have the smallest average bend as well as the lowest bend range with $3.50\degree \pm 1.79\degree$ for lengths $L=3$.
The average bend  increases quadratically with $\Delta$ growing more off-center to
values of $7.52\degree	\pm 4.37\degree$ for pl-s and $6.83\degree\pm	4.13\degree$ for pl-as at $\Delta = 0.2$ and $L=3$.

In all systems, the bend range reduces with the chain length and for extreme $\Delta$-values ($i.e.$, $\Delta=0.2$ and $\Delta=0.8$) the average bend decreases as well. 
For $L=6$ in pl-center the average bend is $3.43\degree	\pm 1.12\degree$ and in pl-s at $\Delta=0.2$ the average bend is $7.18 \degree\pm 2.64 \degree$.

For pl-as, the average bend is asymmetric with respect to $\Delta=0.5$.
Pl-as chains at $\Delta = 0.2$ are slightly less bent ($6.83\degree\pm 4.13\degree$) than at $\Delta = 0.8$ ($7.55\degree \pm 4.37\degree$).
This asymmetry becomes stronger with increasing chain length, as for $L=6$ the average bend at $\Delta=0.2$ is $ 5.17\degree \pm 1.84\degree$ while it is $ 8.17 \degree \pm 2.81 \degree$ for $\Delta = 0.8$.
The asymmetry in bend and bend range might stem from the asymmetry in the bend range between two particles with respect to $\Delta$, where for pl-as$_{\Delta=0.2}$ the bend range is $\Delta\omega_{s} = [-29.32\degree, 29.37\degree]$, and for pl-as$_{\Delta=0.8}$ $\Delta\omega_{as} = [-29.97\degree,29.90\degree]$.
It is interesting to note, that although pl-as$_{\Delta=0.8}$
has a wider bend range than pl-as$_{\Delta=0.2}$, in total the bonding entropy is higher for pl-as$_{\Delta=0.2}$ (expressed in the volume of states in  Fig.~\ref{fig:nstates}) and Fig.~\ref{fig:volume_states}.

\section{ma/mo-systems}
\label{appendix:mamo}
\setcounter{figure}{0}    


\begin{figure*}
\begin{center}
\includegraphics[width=0.8\textwidth]{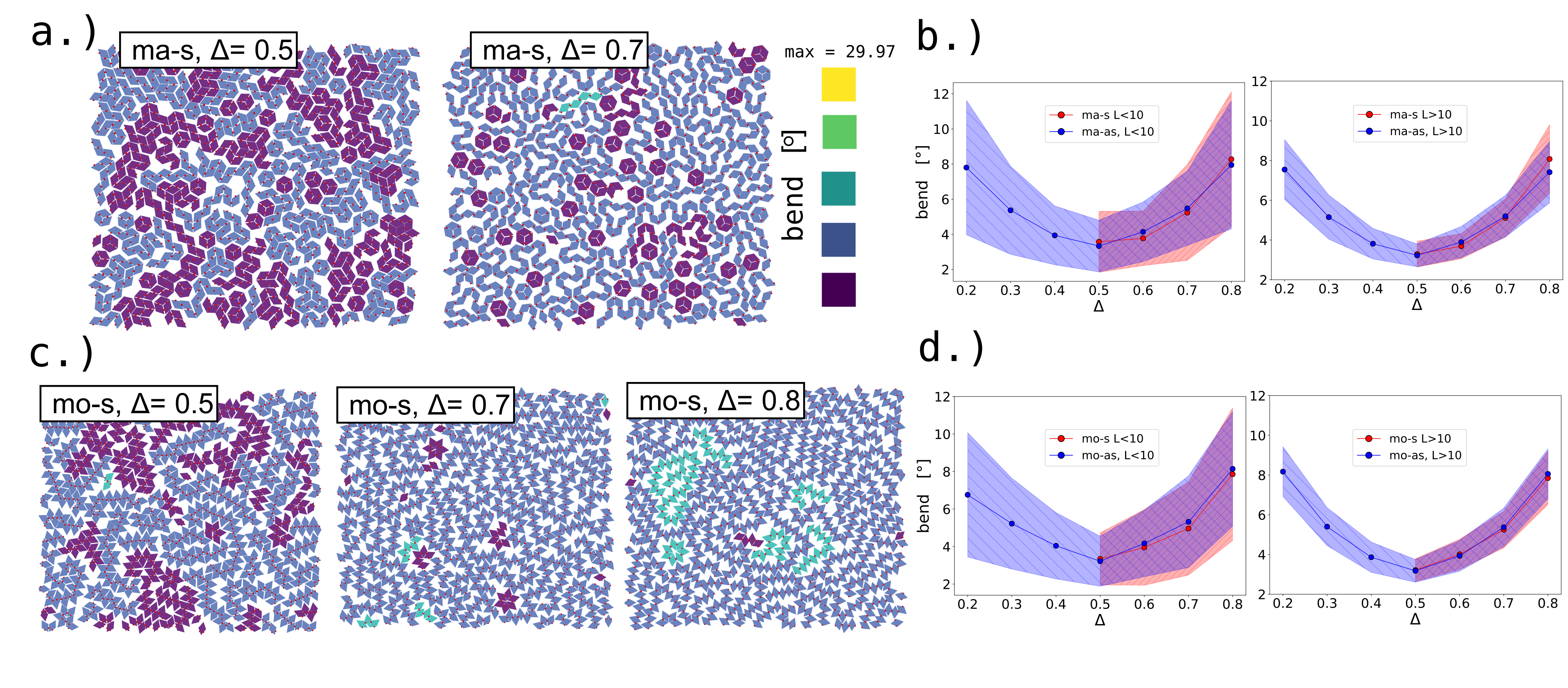}
\caption{\textbf{a.)} Snapshots of ma-s systems at $\epsilon = - 10.2 k_{B}T$ for $\Delta = 0.5$ and $\Delta = 0.7$. The system consist of boxes, chains and loops. Clusters are colored according to their respective bend with a color map that is perceptionally uniform and normalized with the maximum bend observed ($i.e.$, $29.97 \degree$. \textbf{b.)} The average bend of ma-systems and its bend range as function of $\Delta$. \textbf{c.)} Snapshots of mo-s systems at $\epsilon = - 10.2 k_{B}T$ for $\Delta = 0.5$ and $\Delta = 0.7$. The system consists of stars, chains and loops. Clusters are colored according to their respective bend with a perceptionally uniform color map normalized with the maximum bend observed ($i.e.$, $29.97 \degree$).
\textbf{d.)} The average bend of mo-systems and its bend range as function of $\Delta$.
}
\label{fig:moma_flexibility}
\end{center} 
\end{figure*}


\begin{figure}
\begin{center}
\includegraphics[width=0.8\columnwidth]{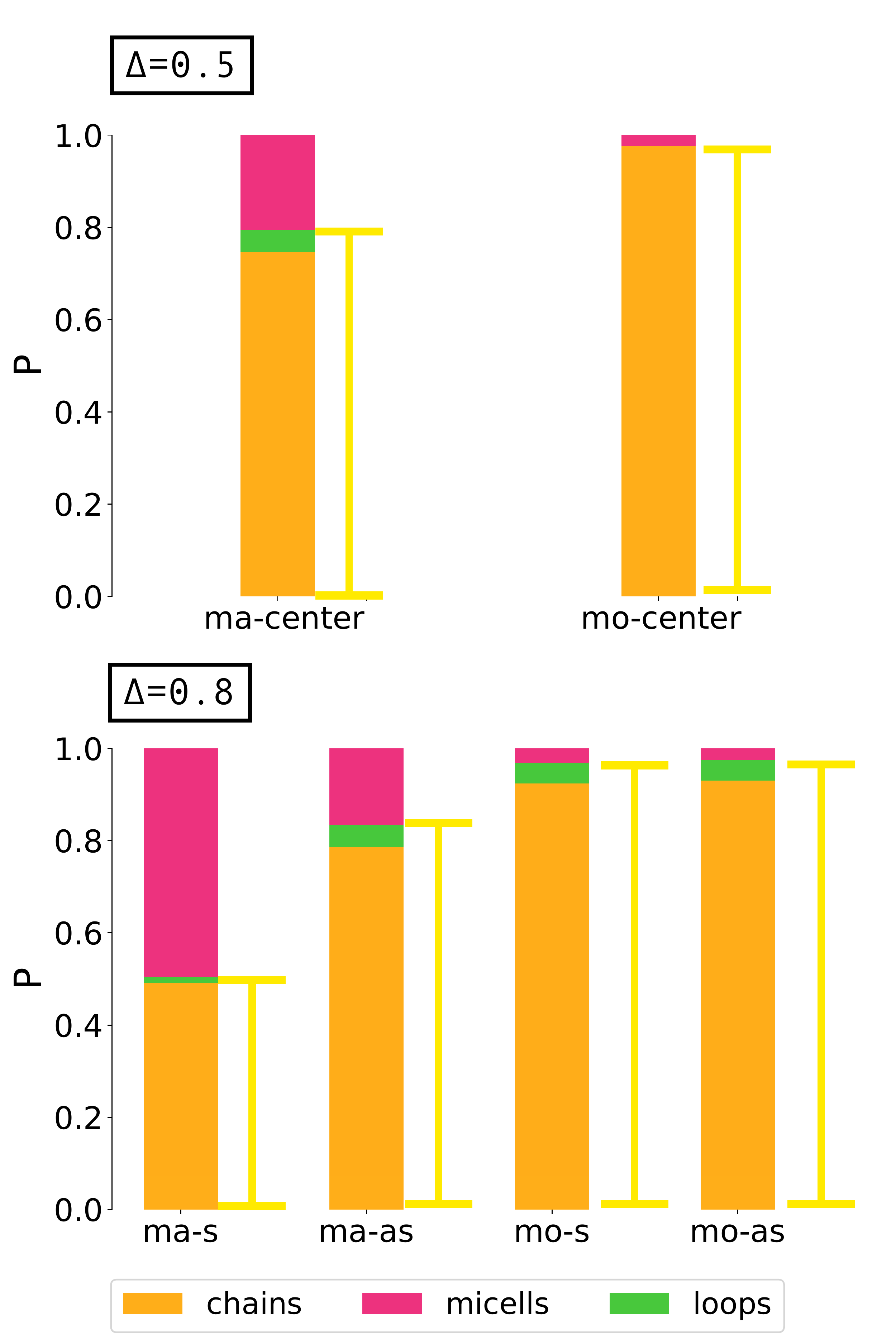}
\caption{Histograms for the yield of chains, (orange) loops (green) and micelles (minimal np-loops, pink) in  manta (ma) and mouse (mo) systems with symmetric (s) and asymmetric (as) topologies and for $\Delta = 0.5$ (top) and $\Delta=0.7$ (bottom). Note that the yellow bars correspond to the sum of chains and loops.}
\label{fig:loops}
\end{center} 
\end{figure}

This section introduces barycentric coordinate systems and discusses the bond flexibility and the cluster types occurring in manta (ma) and mouse (mo) systems. 

\subsection{Barycentric coordinate systems}.
Barycentric coordinate systems are used to visualize mixtures of three components $p_{l}, p_{m}, p_{c}$ with $p_{l} + p_{m} + p_{c} = 1$ as is given in this case.
The percentages $(p_{l},p_{m},p_{c})$ are mapped onto a equilateral triangle. Each triangle edge point represents the extreme in which one $p_{i} = 1$, while the others are 0. 
Mixtures of the three components are Cartesian points $q_{t}$, inside the triangle. Assuming the Cartesian triangle edge points are $x_{l} = (0,0)$, $x_{c} = (1,0)$ and $x_{m} = (1/2, \sqrt{3}/2)$, then $q_{t}$ of a specific mixture $(p_{l},p_{m},p{c})$ is given as 
\begin{equation}
q_{t} = (\frac{1}{2}\cdot (2 p_{c} + p_{m}), \frac{\sqrt(3)}{2} \cdot p_{m})
\end{equation}
We discretize the equilateral triangle into 9 equilateral triangles, hence whenever one cluster type has a yield higher than $2/3$, the heat map takes the dominant cluster type color of the respective triangle edge (blue, yellow, pink). If the system is more mixed, the respective mixed color within in the barycentric triangle is taken on.

\subsection{Bond flexibility}
In general, the average bend and the bend range are the lowest for chains of the center topologies and get higher the more off-center $\Delta$ becomes (see Fig.~\ref{fig:moma_flexibility}). In ma-center for chains with $L<10$ the average bend and bend range are $3.57\degree \pm 1.74 \degree$, whereas for ma-s$_{\Delta=0.8}$ we get $8.276\degree \pm 3.85\degree $ and for ma-as$_{\Delta=0.8}$ we get $7.95\degree \pm 3.64\degree$.
For mo-systems, we observe a similar effect, with an average bend and bend range of $3.34 \degree \pm 1.391\degree$ for mo-center, versus $7.86 \degree \pm 3.54 \degree$ for mo-s$_{\Delta=0.8}$ and $8.14 \degree \pm 3.06\degree$ for mo-as$_{\Delta=0.8}$. 
As the chains become longer, the average bend remains comparable, while the bend range becomes significantly smaller. For chain lengths $L<10$, the average bend and bend range in ma-center are $3.57\degree \pm 1.74 \degree$, while for $L>10$ we get $3.21 \degree\pm 0.57 \degree$.

\subsection{Cluster types}
In contrast to pl-systems, ma- and mo-systems can, besides chains, yield micelles and loops. Micelles are minimal loops, which are 3-np loops for ma-systems (boxes) and 5-np or 6-np loops for mo-systems (stars). 
Fig.~\ref{fig:loops} shows yields - $i.e.$, the percentage
of particles in a particular cluster type - for chains, micelles and loops at $\Delta=0.5$ and $\Delta=0.7$.
We find that for all displayed systems, chains are the most prevalent cluster type, while loops and micelles are observed much less frequently. 
In ma-systems, boxes have a yield of over $0.1$ and loops occur at a yield of $\approx 0.05$, for all displayed $\Delta$-values and symmetric (s) as well as asymmetric (as) topologies. 
In mo-systems, loops and stars are much less frequent with a yield of less than $0.05$. For $\Delta=0.5$, loops completely vanish in both s- and as-topologies.


\footnotesize{
\bibliography{references} 
\bibliographystyle{rsc} 
}
\end{document}